%% file: output.tex
\documentclass[lettersize,journal]{IEEEtran}
\usepackage{amsmath,amsfonts}
\usepackage{algorithmic}
\usepackage{algorithm}
\usepackage{array}
\usepackage[caption=false,font=normalsize,labelfont=sf,textfont=sf]{subfig}
\usepackage{textcomp}
\usepackage{stfloats}
\usepackage{url}
\usepackage{verbatim}
\usepackage{graphicx}
\usepackage{cite}

\usepackage{multirow}
\usepackage{overpic}
\usepackage{bbding}
\usepackage{booktabs}
\usepackage{xcolor}

\begin{document}

\title{RAWIW: RAW Image Watermarking Robust to ISP Pipeline}

\author{Kang Fu, Xiaohong Liu, Jun Jia, Zicheng Zhang, Yicong Peng,\\ Jia Wang, and Guangtao Zhai, \emph{Senior Member, IEEE} % <-this % stops a space
\IEEEcompsocitemizethanks{\IEEEcompsocthanksitem Kang Fu, Jun Jia, Zicheng Zhang, Yicong Peng, Jia Wang, and Guangtao Zhai are with the Institute of Image Communication and Network Engineering Shanghai Jiao Tong University, 200240 Shanghai, China. E-mail:\{fuk20-20, jiajun0302, zzc1998, jack-sparrow, jiawang, zhaiguangtao\}@sjtu.edu.cn.\protect  \quad Xiaohong Liu is with John Hopcroft Center, Shanghai Jiao Tong University, Shanghai 200240, China. E-mail: xiaohongliu@sjtu.edu.cn.}
        % <-this % stops a space
}% <-this % stops a space

\maketitle

\begin{abstract}
Invisible image watermarking is essential for image copyright protection. Compared to RGB images, RAW format images use a higher dynamic range to capture the radiometric characteristics of the camera sensor, providing greater flexibility in post-processing and retouching. Similar to the master recording in the music industry, RAW images are considered the original format for distribution and image production, thus requiring copyright protection. Existing watermarking methods typically target RGB images, leaving a gap for RAW images. To address this issue, we propose the first deep learning-based \underline{RAW} \underline{I}mage \underline{W}atermarking (RAWIW) framework for copyright protection. Unlike RGB image watermarking, our method achieves cross-domain copyright protection. We directly embed copyright information into RAW images, which can be later extracted from the corresponding RGB images generated by different post-processing methods. To achieve end-to-end training of the framework, we integrate a neural network that simulates the ISP pipeline to handle the RAW-to-RGB conversion process. To further validate the generalization of our framework to traditional ISP pipelines and its robustness to transmission distortion, we adopt a distortion network. This network simulates various types of noises introduced during the traditional ISP pipeline and transmission. Furthermore, we employ a three-stage training strategy to strike a balance between robustness and concealment of watermarking. Our extensive experiments demonstrate that RAWIW successfully achieves cross-domain copyright protection for RAW images while maintaining their visual quality and robustness to ISP pipeline distortions.
\end{abstract}


% \section{Introduction}
% \IEEEPARstart{T}{his} file is intended to serve as a ``sample article file''
% for IEEE journal papers produced under \LaTeX\ using
% IEEEtran.cls version 1.8b and later. The most common elements are covered in the simplified and updated instructions in ``New\_IEEEtran\_how-to.pdf''. For less common elements you can refer back to the original ``IEEEtran\_HOWTO.pdf''. It is assumed that the reader has a basic working knowledge of \LaTeX. Those who are new to \LaTeX \ are encouraged to read Tobias Oetiker's ``The Not So Short Introduction to \LaTeX ,'' available at: \url{http://tug.ctan.org/info/lshort/english/lshort.pdf} which provides an overview of working with \LaTeX.
\input{Contents/Introduction}

\input{Contents/Related_Works}

\input{Contents/Method}

\input{Contents/Experiments}

\input{Contents/Limitations}

\input{Contents/Conclusion}

\bibliographystyle{IEEEtran}
\bibliography{output} 
% \bibitem{ref1}
% {\it{Mathematics Into Type}}. American Mathematical Society. [Online]. Available: https://www.ams.org/arc/styleguide/mit-2.pdf

% \bibitem{ref2}
% T. W. Chaundy, P. R. Barrett and C. Batey, {\it{The Printing of Mathematics}}. London, U.K., Oxford Univ. Press, 1954.

% \bibitem{ref3}
% F. Mittelbach and M. Goossens, {\it{The \LaTeX Companion}}, 2nd ed. Boston, MA, USA: Pearson, 2004.

% \bibitem{ref4}
% G. Gr\"atzer, {\it{More Math Into LaTeX}}, New York, NY, USA: Springer, 2007.

% \bibitem{ref5}M. Letourneau and J. W. Sharp, {\it{AMS-StyleGuide-online.pdf,}} American Mathematical Society, Providence, RI, USA, [Online]. Available: http://www.ams.org/arc/styleguide/index.html

% \bibitem{ref6}
% H. Sira-Ramirez, ``On the sliding mode control of nonlinear systems,'' \textit{Syst. Control Lett.}, vol. 19, pp. 303--312, 1992.

% \bibitem{ref7}
% A. Levant, ``Exact differentiation of signals with unbounded higher derivatives,''  in \textit{Proc. 45th IEEE Conf. Decis.
% Control}, San Diego, CA, USA, 2006, pp. 5585--5590. DOI: 10.1109/CDC.2006.377165.

% \bibitem{ref8}
% M. Fliess, C. Join, and H. Sira-Ramirez, ``Non-linear estimation is easy,'' \textit{Int. J. Model., Ident. Control}, vol. 4, no. 1, pp. 12--27, 2008.

% \bibitem{ref9}
% R. Ortega, A. Astolfi, G. Bastin, and H. Rodriguez, ``Stabilization of food-chain systems using a port-controlled Hamiltonian description,'' in \textit{Proc. Amer. Control Conf.}, Chicago, IL, USA,
% 2000, pp. 2245--2249.

\newpage

 

% \end{IEEEbiography}



\vfill

\end{document}

%% file: Contents/Introduction.tex
\section{Introduction}

\begin{figure*}[!htbp]
  \centering
  \includegraphics[width=0.95\linewidth]{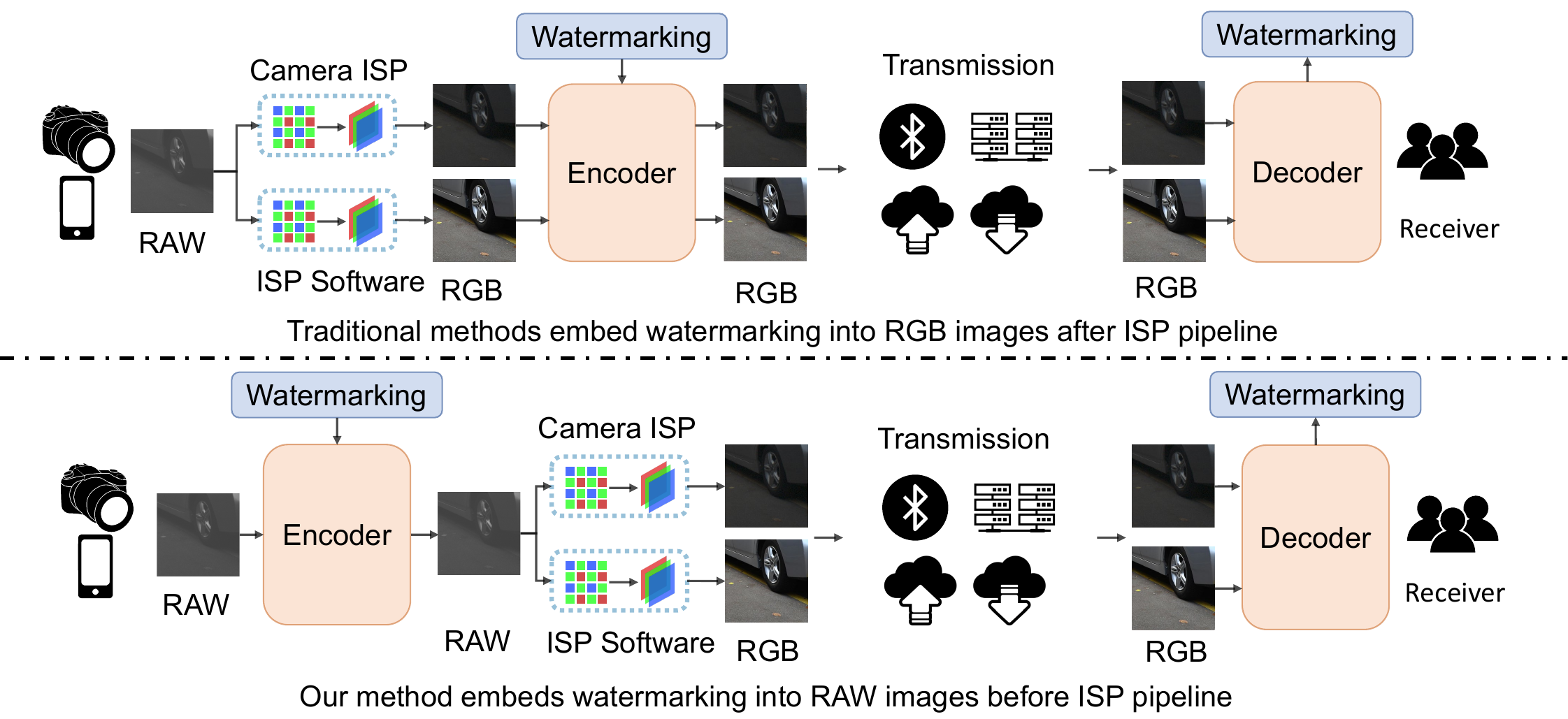}
  \caption{Schematic diagram of our method compared with previous methods. Our method encodes the watermarking messages into RAW images before the ISP pipeline and decodes the generated RGB images to extract the watermarking. This approach allows the same RAW image to be used with different ISP pipelines, producing varieties of RGB images with different colors and qualities with the extractable watermarking. }
  \label{figintro}
\end{figure*}

\IEEEPARstart{I}{n} the era of information, safeguarding intellectual property is vital, especially for multimedia resources on the Internet, such as recordings, photographs, and videos. Among these, photographs are widespread and require significant copyright protection. Invisible image watermarking plays a crucial role in the field of information security by securing the copyrights of digital images and tracing unauthorized disclosures. This technique involves creating a watermarked image that appears similar to the original image but contains unique copyright information. At the same time, the hidden watermarking must be robust enough to withstand noises and distortions that may occur during transmission, such as Gaussian noise and JPEG compression. A reliable watermarking algorithm should ensure that the hidden information remains detectable even after encountering these challenges. Traditionally, existing watermarking algorithms have focused on manipulating RGB images and hiding information in spatial or transform domains. However, professional photographers nowadays prefer working with RAW images to achieve more satisfactory results, as RAW images offer greater flexibility and ease of manipulation. Unlike RGB images, RAW images store 10-16 bits of unprocessed scene radiance, capturing a higher dynamic range and providing more room for post-processing. The popularity of smartphones like Huawei P20, iPhone 13 Pro, and Samsung Galaxy S22, which support capturing RAW images, indicates a potential surge in the number of RAW images in the near future. Moreover, RAW images are akin to works of art and should not be distributed without proper authorization. Publishers often receive RAW images and use them to generate different RGB images. As a result, applying watermarking algorithms to RAW images is critical for copyright protection but has not received sufficient attention in previous works. In conclusion, with the increasing prevalence of multimedia resources, particularly photographs, on the Internet, the importance of protecting intellectual property through invisible image watermarking cannot be overstated. Paying attention to RAW images and developing effective watermarking techniques for them is crucial in ensuring copyright protection in this digital age. 

Unlike RGB images, the copyright protection of RAW images faces significant challenges. Applying watermarking directly to RGB images poses the risk of leaking the original RAW image, and the leaked RAW image can generate unwatermarked RGB images, thereby failing to achieve the goal of copyright protection. On the other hand, attempting to decode the watermarking from RAW images presents difficulties since RAW images are not readily accessible in most cases. The existing invisible image watermarking methods typically hide and extract information within the same image format, specifically in RGB color space. If we were to apply these existing methods directly to RAW images, extracting the watermarking from RAW images would be impractical due to limited accessibility. Additionally, a RAW image can be transformed into different RGB images by various Image Signal Processing (ISP) pipelines. Hence, the process of RAW image watermarking involves encoding the watermarking into a RAW image and subsequently decoding the watermarking from the corresponding RGB images after undergoing various ISP pipelines. This multistage process aims to ensure copyright protection while accounting for the different transformations that the RAW image undergoes during image processing. In summary, protecting the copyright of RAW images requires overcoming various challenges, such as preventing leakage of the original RAW image and handling the transformation of RAW images to different RGB images by various ISP pipelines. Existing watermarking methods designed for RGB images are not directly applicable to RAW images due to the differences in accessibility and image format, emphasizing the need for specialized techniques for RAW image watermarking.

To address the aforementioned challenges, Meerwald \textit{et al.}\cite{meerwald2009watermarking} proposed frequency domain transform based RAW image watermarking. This paper introduces the pioneering deep learning watermarking-based copyright protection framework for RAW images, called \underline{RAW} \underline{I}mage \underline{W}atermarking (RAWIW). RAWIW utilizes Convolutional Neural Networks (CNN) for embedding and detecting the watermarking. The comparison between our RAWIW and the RGB watermarking method is illustrated in Fig \ref{figintro}. As discussed earlier, our proposed RAWIW encodes the watermarking information into the RAW image before applying the Image Signal Processing (ISP) pipeline. This approach enhances encoding efficiency by applying watermarking to the RAW image to undergo different retouching methods (\textit{ISP pipelines}), producing RGB images with diverse color hues and tones. The watermark decoder can then extract the watermarking information from these RGB images, irrespective of the employed retouching. In contrast, RGB watermarking necessitates encoding the watermarking information into each RGB image generated by distinct retouching methods. This process is time-consuming and computationally intensive, as each RGB image must be processed individually. By embedding the watermarking into RAW images before applying ISP pipelines, our framework streamlines the encoding process and enhances encoding efficiency. Furthermore, our method enhances decoding robustness by allowing the decoder to adapt to different retouching methods through flexible changes during the training pipeline. Consequently, our proposed method provides an effective solution for safeguarding the copyright of RAW images.

The RAWIW framework comprises five essential modules: encoder, decoder, discriminator, distortion network, and deep ISP pipeline. Within this framework, the encoder embeds watermarking information into a RAW image, while the decoder retrieves the hidden information from the corresponding RGB image. To ensure the encoded information's robustness during the transfer from RAW to RGB format, a deep differentiable ISP pipeline~\cite{ignatov2020aim,ronneberger2015u,dai2020awnet} is integrated, simulating camera image processing. Additionally, we incorporate a distortion network after the ISP module to simulate the distortions encountered during transmission, further enhancing the robustness of the watermarking against these distortions. The discriminator and encoder modules are trained adversarially to improve the concealment of the watermarking. Moreover, we employ an effective three-stage training strategy to strike a balance between the robustness and concealment of the watermarking. The framework has undergone evaluations using two datasets of RAW images, demonstrating outstanding performance in terms of decoding accuracy and visual quality. Overall, this paper presents a comprehensive approach for adding watermarking to RAW images, applicable in various scenarios, such as digital rights management and copyright protection. This robust and efficient framework can significantly contribute to safeguarding the copyright of RAW images in diverse applications.

The contributions of this paper are threefold:
\begin{itemize}

\item
{
To the best of our knowledge, this paper presents the \textbf{first deep learning-based cross-domain RAW image watermarking method named RAWIW} which can encode watermarking to RAW images while decoding watermarking from the corresponding retouched RGB images. This method can achieve the protection of copyright and ownership for RAW images.
}

\item {
We are \textbf{the first to propose a RAW image encoder that considers Bayer patterns and a distortion network that simulates the gap of different ISP pipelines and the distortion of transmission} while we use an effective three-stage training strategy for our method, which can achieve a good trade-off between robustness and concealment of watermarking.
}

\item{
Extensive experiments demonstrate that the proposed method has good concealment while being robust to different ISP pipelines.
}

\end{itemize}

The paper is structured as follows. Section \ref{sec:Related Work} provides an overview of the background and related work related to the proposed method. In Section \ref{sec:Method}, we elaborate on our proposed watermarking method in detail. Next, Section \ref{sec:Experiments} outlines the experimental setup and showcases the results obtained from our experiments. In Section \ref{sec:Limitations} and Section \ref{sec:Conclusion}, we discuss the advantages and limitations of the proposed method.

%% file: Contents/Related_Works.tex
\section{Related Work}
\label{sec:Related Work}
\subsection{Invisible Information Hiding}
Invisible information hiding can be broadly classified into two main categories: steganography and digital watermarking. Steganography finds widespread use in the field of information security, with its primary objective being to ensure that information is accessible only to the intended recipient while remaining concealed from unauthorized individuals. Steganography can be further divided into two classes: spatial domain and transfer domain. The classical spatial domain method is Least Significant Bit (LSB), where the hidden message replaces the least significant bits of the cover image. However, this approach alters the statistical properties of the cover image, making it easily detectable by steganalysis. As a result, simple LSB steganography is ineffective in practical applications. On the other hand, steganography in the transfer domain~\cite{marvel1999spread,johnson1998exploring,wang2016rate, lu2020secure, li2019jpeg, li2019shortening, zhang2016decomposing} leverages the statistical characteristics of the image to conceal information. In recent years, numerous steganographic techniques based on CNN have been introduced. The impressive non-linear fitting capability of CNN allows for embedding and extracting information without the need for intricate manual feature extractions. Some well-established methods, such as SSGAN~\cite{shi2018ssgan} and ASDL-GAN\cite{tang2017automatic}, have proposed modifications to the redundant information of the cover image to hide the desired information. \cite{guo2023hierarchical,liu2022pscc} use CNN to realize image forgery detection and localization.

Digital watermarking is a crucial aspect of information hiding, involving the insertion of concise messages into images to protect copyright and assert authorship. Unlike steganography, digital watermarking requires a high level of robustness against transmission distortion. Traditional digital watermarking methods can be classified into two categories: spatial~\cite{karybali2006efficient, pereira2001optimal, kim2003invariant, yang2021high} and transfer~\cite{6486549,birney1995modeling, hernandez2000dct, cheng2003robust, barni2001new, zheng2003rst, xiang2008invariant, wang2023udtcwt, huang2023robust} domain approaches.

Zhu \textit{et al.}\cite{Zhu_2018_ECCV} presented an innovative approach for achieving robust image watermarking through adversarial learning. Their pioneering work demonstrated robustness against various distortions, including Gaussian blurring, pixelwise dropout, cropping, and JPEG compression. Building upon Zhu's framework, Tancik \textit{et al.}\cite{tancik2020stegastamp} introduced Stegastamp, which incorporated shooting noise to achieve robust watermarking for shooting screen and printed images. Jia \textit{et al.}~\cite{jia2020rihoop} proposed RIHOOP, utilizing differentiable 3-D rendering operations to simulate distortions resulting from camera imaging. However, unlike the aforementioned methods that embed the watermarking in RGB format, our method embeds the watermarking in RAW format and extracts it from RGB format. This distinction enables us to address specific challenges related to RAW images and attain effective copyright protection in diverse scenarios.

\subsection{Image Signal Processing Pipeline}
The Image Signal Processing (ISP) pipeline in a camera is utilized to transform RAW images captured by the camera sensor into RGB images that are perceptually optimized for the Human Visual System (HVS). To achieve exceptional visual quality, the Camera ISP pipeline consists of various modules, such as demosaicing, white balance, color correction, color mapping, gamma correction, image enhancement, noise reduction, and sharpening. However, many of these sub-modules are non-differentiable, meaning that they do not allow for the backpropagation of gradients through them. This non-differentiability poses a challenge when attempting to train the complete ISP pipeline end-to-end in a neural network, hindering the optimization process. As a result, effectively incorporating the ISP pipeline into the neural network architecture requires specialized techniques to overcome these non-differentiable components and ensure smooth training and optimization.

In contrast to the traditional ISP pipeline, where each sub-module is treated separately, the deep ISP pipeline operates on RAW images to produce RGB images using a deep neural network. Recent methodologies~\cite{C5,8259342,10.1145/2980179.2982399} based on CNN have demonstrated remarkable advancements in various ISP tasks, showcasing the superiority of CNN in this domain. Consequently, using a CNN instead of the entire ISP pipeline is feasible. Many efforts have been made in recent years to train deep networks to learn the ISP pipeline. Schwartz \textit{et al.}\cite{8478390} created a dataset containing RAW images and their corresponding RGB images and proposed the DeepISP model. This model establishes a mapping between RAW low-light images and well-lit processed RGB images. CameraNet\cite{9329084} comprises two distinct CNN modules designed to address two sets of relatively uncorrelated subtasks in an ISP pipeline: restoration and enhancement. Ignatov \textit{et al.}\cite{ignatov2020replacing} introduced an inverted pyramidal architecture named PyNET, capable of processing images at five distinct levels, to learn a diverse set of features at each level. They also collected a dataset containing paired RAW and RGB images, which was subsequently utilized in two challenges\cite{9022218,ignatov2020aim}. The top-performing methods in these challenges were MW-ISPNet~\cite{ignatov2020aim} and AWNet~\cite{dai2020awnet}, both using a Discrete Wavelet Transform (DWT)-based decomposition to replace upsampling and downsampling operations. MW-ISPNet integrates MWCNN~\cite{8575273} with RCAN~cite{zhang2018rcan} models, while AWNet employs an attention mechanism. Zhang \textit{et al.}~\cite{RAW-to-sRGB} introduced a light ISP network that builds upon the MW-ISPNet architecture and incorporates image alignment during training. This image alignment has led to the current state-of-the-art performance of the light ISP network.

The proposed RAWIW framework incorporates a deep ISP pipeline that represents the traditional ISP pipeline in a differentiable manner. This deep ISP pipeline is constructed using CNN, enabling end-to-end training of the complete framework.By utilizing a deep ISP pipeline, the proposed framework achieves a superior balance between accuracy and computational efficiency. The differentiable nature of the deep ISP pipeline enables efficient backpropagation of gradients throughout the entire framework, facilitating the optimization process and enhancing training effectiveness. In summary, the adoption of a deep ISP pipeline in the RAWIW framework not only enables end-to-end training but also improves the overall performance by efficiently managing computational resources and optimizing the training process.

%% file: Contents/Method.tex
\section{Method}
\label{sec:Method}
\begin{figure*}[!htbp]
  \centering
  \includegraphics[width=\linewidth]{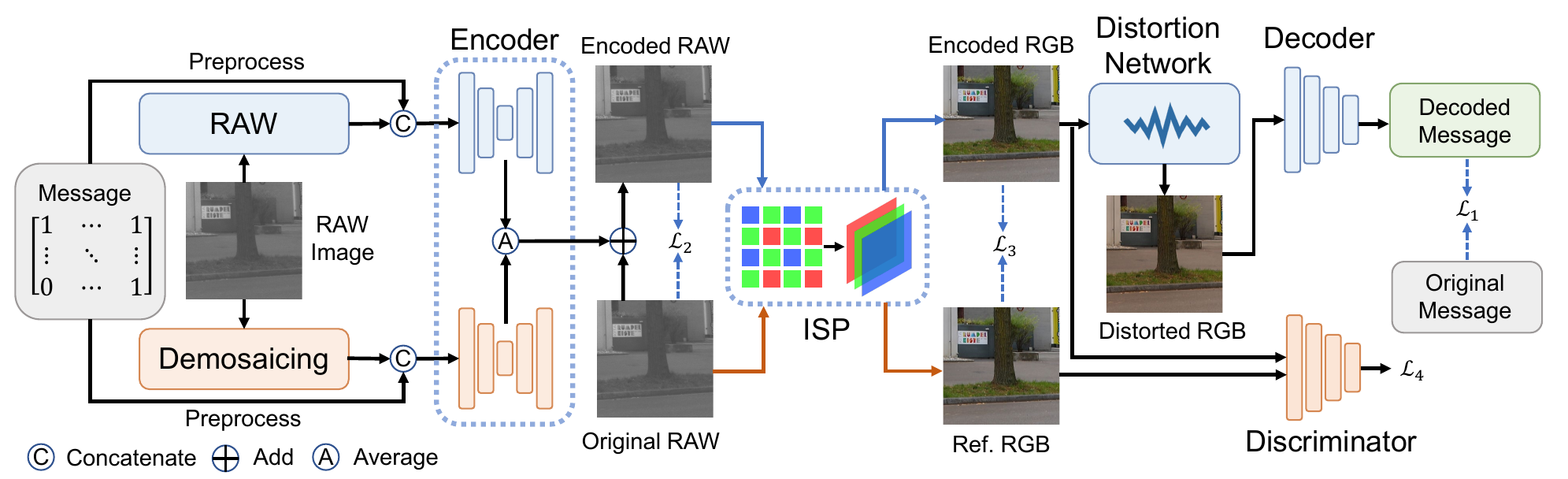}
  \caption{Overall architecture of the RAWIW framework. RAWIW consists of three major modules: (1) the Encoder, Decoder and Discriminator forms the watermarking module, which embeds the watermarking message into a given RAW and decodes it from the converted RGB image. (2) the deep ISP module simulates the traditional ISP pipeline while keeping the process differentiable. (3) the Distortion Network adds noise and compression to the encoded RGB image, which simulates distortions introduced during the traditional ISP pipeline as well as transmission. Losses $\mathcal{L}_{1 \sim 4}$ are: watermarking message loss, RAW image loss, converted RGB image loss, and discriminator loss respectively.}
  \label{section:framework}
\end{figure*}

In this section, we introduce the proposed RAW image watermarking method named RAWIW. 
The schematic overview of the proposed method, as depicted in Fig \ref{section:framework}, comprises a watermarking module, a deep ISP module, and a distortion network. The watermarking module aims at embedding and extracting the watermarking, while the deep ISP module converts RAW images to RGB images. The distortion network introduces simulated noises into the framework, thereby enhancing the robustness of the watermarking against a range of adversarial attacks. The following subsections will describe these modules respectively in detail.

\subsection{Problem Formulation}
Our proposed methods embed a copyright message $\mathbf{M}$ within a RAW image $\mathbf{R}$ via a RAW image encoder $\mathbf{E}$ with weights $\mathbf{\theta_{E}}$ to produce an encoded RAW image $\mathbf{R_{e}}$. The ISP pipeline $\mathbf{P}$ is subsequently utilized to convert the encoded RAW image $\mathbf{R_{e}}$ to an RGB image $\mathbf{I_{e}}$ and also convert the original RAW image $\mathbf{R}$ to  RGB image $\mathbf{I_{o}}$. Finally, our proposed method utilizes an RGB image decoder $\mathbf{D}$ with weights $\mathbf{\theta_{D}}$ to extract the embedded message $\mathbf{M_{d}}$ from the RGB image $\mathbf{I_{e}}$. The above process can be expressed by the following four formulas:
\begin{equation}
    \mathbf{R_{e}} = \mathbf{E}_{\theta_{E}}(\mathbf{R}, \mathbf{M}), \mathbf{I_{e}} = \mathbf{P}(\mathbf{R_{e}}), \mathbf{I_{o}} = \mathbf{P}(\mathbf{R}), \mathbf{M_{d}} = \mathbf{D}_{\theta_{D}}(\mathbf{I_{e}}).
\end{equation}
For the purpose of the copyright protection for RAW images, it is imperative to achieve the congruence between $\mathbf{M_{d}}$ and $\mathbf{M}$, while simultaneously ensuring that the visual quality of both images remain uncompromised, Therefore, we need to optimize the parameters during the training process:
\begin{equation}
    \arg \min_{\theta_{E} , \theta_{D}} d_{rgb}(\mathbf{I_{o}}, \mathbf{I_{e}}) + d_{raw}(\mathbf{R}, \mathbf{R_{e}}) + d_{bin}(\mathbf{M}, \mathbf{M_{d}}),
\end{equation}
where $d_{rgb}$ stands for the RGB image distance, and analogously for $d_{raw}$, $d_{bin}$.

\subsection{Image Watermarking}
The image watermarking module is composed of three parts: encoder, decoder, and discriminator. The encoder is responsible for embedding the watermarking information into the RAW image, while the decoder extracts the watermarking information from the corresponding RGB image. The discriminator and encoder modules are trained adversarially to enhance the concealment of the watermarking information.

\begin{figure*}[!htbp]
  \centering
    \renewcommand{\tabcolsep}{1.0pt} %
    \renewcommand{\arraystretch}{0.8}
    \begin{tabular}{cccccc}
        \includegraphics[width=0.16\textwidth]{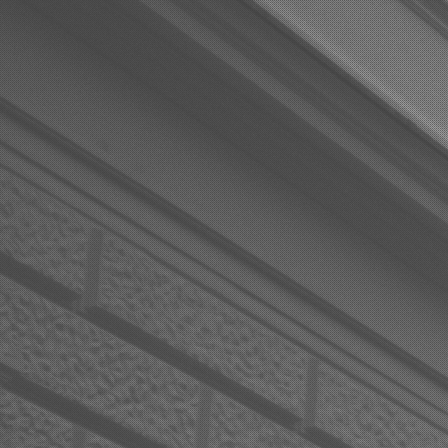} & 
        \includegraphics[width=0.16\textwidth]{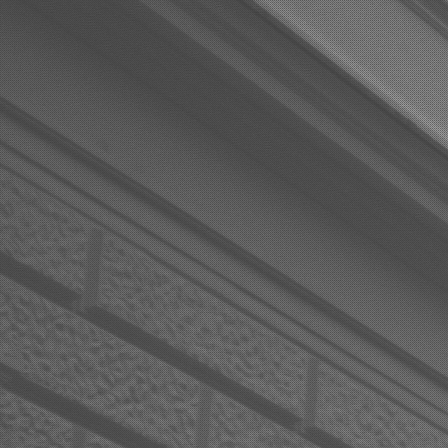} & 
        \includegraphics[width=0.16\textwidth]{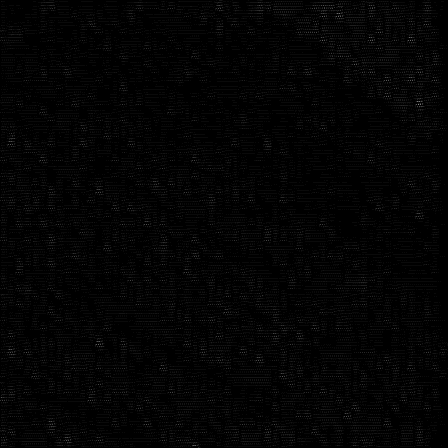} & 
        \includegraphics[width=0.16\textwidth]{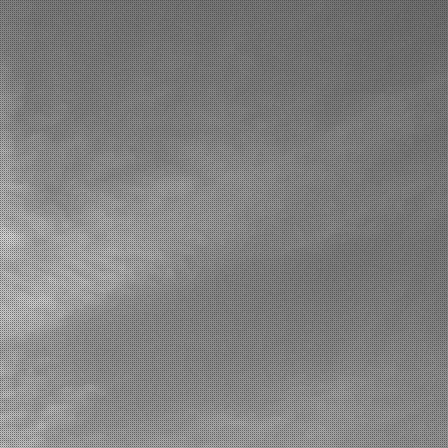}&
        \includegraphics[width=0.16\textwidth]{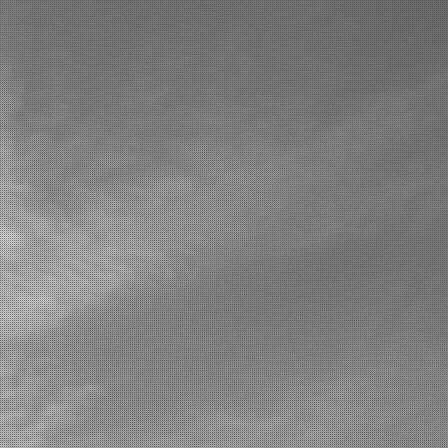}&
        \includegraphics[width=0.16\textwidth]{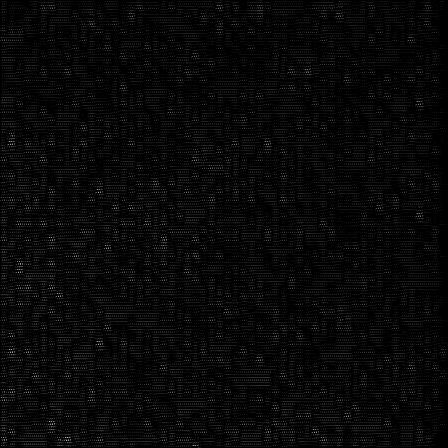}\\ 
        \includegraphics[width=0.16\textwidth]{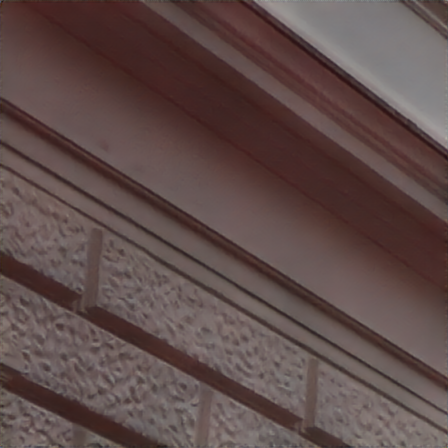} & 
        \includegraphics[width=0.16\textwidth]{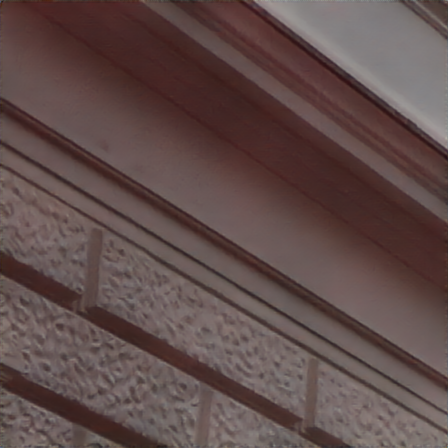} & 
        \includegraphics[width=0.16\textwidth]{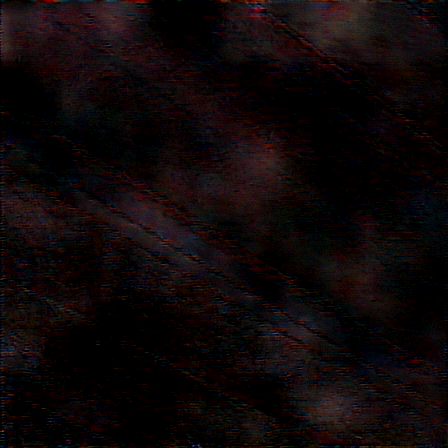}& 
        \includegraphics[width=0.16\textwidth]{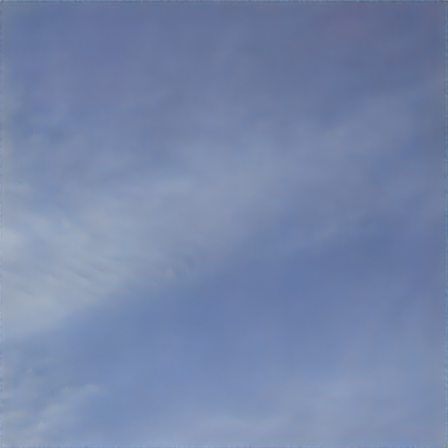}&
        \includegraphics[width=0.16\textwidth]{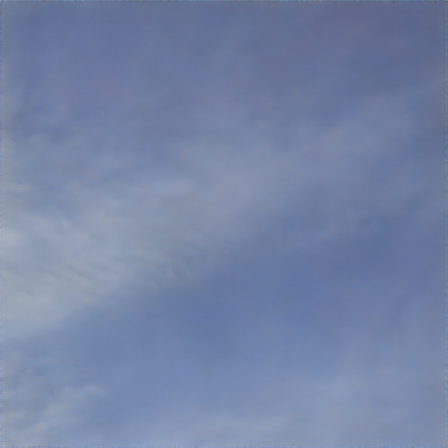}&
        \includegraphics[width=0.16\textwidth]{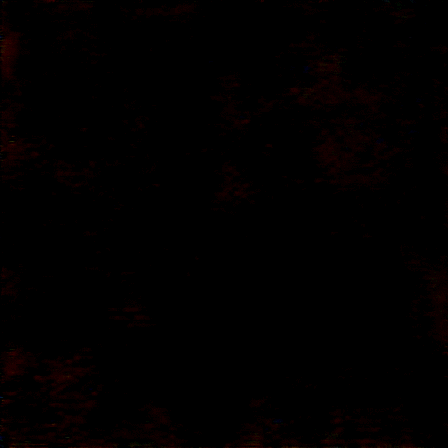} \\
        (a) Original & (b) Encoded & (c) Residual  & (d) Original  & (e) Encoded  & (f) Residual\\
    \end{tabular}
  \caption{Examples of encoded and residual images. The first and second columns are RAW and RGB images separately. The residual image is the absolute difference between the encoded image and the original image (To see the difference clearly, we multiply the RAW residuals and the RGB residuals by 10). These examples have 100 bit encoded messages.}
  \label{figresidual}
\end{figure*}

\textbf{Encoder}: 
The primary function of the encoder is to integrate watermarking information into a RAW image $\mathbf{R}$, producing an encoded RAW image $\mathbf{R_{e}}$ that exhibits the minimal perceptual differentiation compared to $\mathbf{R}$. Furthermore, the RGB images $\mathbf{I_{e}}$ derived from $\mathbf{R_{e}}$ exhibit the minimal perceptual differentiation compared to $\mathbf{I_{o}}$.
The watermarking message $\mathbf{M} \in \{0, 1\}^{L}$ is represented as a binary vector of length $\mathbf{L}$. This binary vector is first processed through a fully connected layer, resulting in a tensor $\mathbf{T} \in \mathbb{R}^{\frac{H}{4} \times \frac{W}{4} \times 1}$. Subsequently, the tensor $\mathbf{T}$ is upsampled to form a tensor $\mathbf{T^{'} } \in \mathbb{R}^{H \times W \times 1 }$, which is then concatenated with the RAW image $\mathbf{R} \in \mathbb{R}^{H \times W \times 1}$. This preprocess of watermarking message benefits the convergence of encoder and decoder\cite{tancik2020stegastamp}.
Considering the properties of bayer pattern in RAW images, we use two U-Net alike networks to compose our encoder, where one is used to process the concatenated tensor $\mathbf{T^{'}} \otimes \mathbf{R}$ of RAW image and watermarking, where $\otimes$ denotes the concatenation operation of tensor. The other is used to concatenated tensor $\mathbf{T^{'}} \otimes \mathbf{R_{d}}$ of demosaicing  RAW image $\mathbf{R_{d}} \in \mathbb{R}^{H \times W \times 4}$ and watermarking. Where demosaicing RAW image $\mathbf{R_{d}} = U( \mathbf{R_{re}} \otimes \mathbf{R_{b}} \otimes \mathbf{R_{gr}} \otimes \mathbf{R_{gb}})$, $\mathbf{R_{re}}$, $\mathbf{R_{b}}$, $\mathbf{R_{gr}}$, $\mathbf{R_{gb}}$ denotes the subimages of red pixels, blue pixels, green pixels in upper left and green pixels in lower right respectively and $U$ means the upsampling 2 times. 
% In subsequent ablation experiments, we also performed ablation experiments using only RAW images, only demosaiced images and both.
Finally, the results from two networks are averaged to get a single channel RAW residual $\mathbf{R_{r}} \in \mathbb{R}^{ H \times W \times 1}$, which will add to the original RAW image to get container RAW image $\mathbf{R_{e}} \in \mathbb{R}^{ H \times W \times 1}$.
We present examples of encoded images and residuals in Fig \ref{figresidual}. In subsequent ablation experiments, we also conduct the experiments that only uses RAW images, only demosaiced images and both. The motivation behind the utilization of the two-streamed encoder lies in our aspiration to incorporate supplementary pixel spatial information into the encoder through the implementation of a demosaicing module. It is crucial to emphasize that similar modules have been developed for existing deep ISP pipelines, indicating their effectiveness and relevance. By leveraging these modules, our objective is to enhance the overall performance and capabilities of our framework, resulting in improved image quality and fidelity of the encoded images. The ablation experiment serves as an illustration, demonstrating that the introduction of additional pixel spatial information can indeed enhance image quality and decoding accuracy.

\textbf{Decoder}: 
The function of the decoder is to extract the watermarking from the RGB image produced by the deep ISP pipeline (image will also go through the distortion network if using it in training). We adopt the same decoder structure as Stegastamp~\cite{tancik2020stegastamp}. To extract the output message from the RGB images, we resort multiple convolutional layers and multiple Multilayer Perceptron (MLP), which are designed to extract relevant features and patterns from the image to recover watermarking. In the end, we use a Sigmoid function to map output values to a probability score between 0 and 1. The length of the output message is determined by the number of units in final MLP, which have the same length as the input message. The decoder network is supervised using the cross-entropy loss between copyright watermarking $\mathbf{M}$ and embeded watermarking $\mathbf{M_{d}}$.

\textbf{Discriminator}: 
Since the adversarial training improves the visual quality of encoded images~\cite{Zhu_2018_ECCV}, we introduce a WGAN-style discriminator~\cite{arjovsky2017wasserstein} in our RAWIW to supervise the training of our encoder. The discriminator network comprises a series of convolutional layers followed by a max pooling. It is used to predict whether a watermarking is encoded in the RGB image which is used as a perceptual loss for the whole framework. The discriminator is trained using an input image and its corresponding encoded image, where the Wasserstein loss function is employed as a supervised signal.

\subsection{Deep ISP Pipeline}
RAWIW requires a module to transform RAW images into RGB images. Given the complexity of the traditional ISP pipeline, which includes some non-differentiable operations, we have chosen to utilize the deep ISP pipeline in our investigation. Specifically, we employ MW-ISPNet~\cite{ignatov2020aim} and AWNet~\cite{dai2020awnet}, as mentioned above. Additionally, we trained a UNet~\cite{ronneberger2015u}-style architecture ISP pipeline for conducting experiments. During the training process of the watermarking module, the parameters of the entire deep ISP pipeline are kept fixed. The reason for not modifying the parameters of the ISP pipeline to enhance the concealment of watermarking is that our primary objective is to enable the encoder to learn how to directly encode information into RAW images and the decoder to learn how to decode information from RGB images. Moreover, modifying the ISP pipeline could introduce additional computational complexity, potentially impacting the overall system performance. To compare the impact of different ISP pipelines on the entire framework, we conducted relevant ablation experiments, and the specific experimental results are presented in Section \ref{section:abst}.

\subsection{Distortion Network}
Since the deep ISP pipeline is trained to establish a mapping between RAW images and RGB images, this mapping can be significantly influenced by the ISP pipeline of data collection devices. However, for RAW image copyright protection, the watermarking must remain robust to traditional ISP pipelines as well. To enhance the decoder's robustness to these distortions (we consider the disparity between the deep ISP pipeline and traditional ISP pipelines as distortion), we introduce a distortion network in front of the decoder during training. In our approach, we address the visual gap between RGB images generated by different ISP pipelines, primarily manifested in differences in color temperature and compression. Additionally, to ensure the watermarking's robustness to transmission noises, we incorporate operations such as differentiable color temperature adjustment, JPEG compression~\cite{zhang2020towards}, Gaussian noise, saturation, contrast, and brightness adjustment into the distortion network. During training, the degrees of these distortions are randomly chosen within a range related to the training epoch, facilitating the decoder's convergence. Fig \ref{figdistortion} visually illustrates the image distortion process. The introduction of the distortion network enhances the decoder's capability to handle various distortions, ensuring robustness and maintaining the effectiveness of the watermarking even when faced with the challenges posed by traditional ISP pipelines. The details of the distortions are as follows:

\begin{figure*}[!htbp]
  \centering
  \begin{overpic}[width=\textwidth]{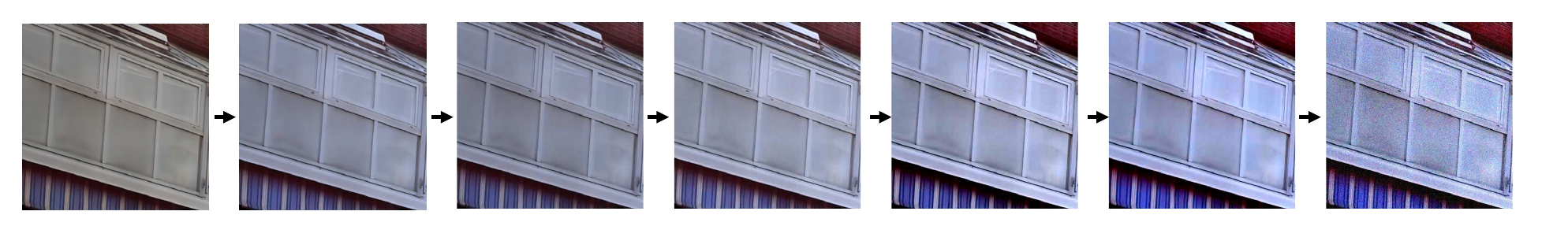}
  \put(4.2,-2.2){(a) Input}
  \put(17.5,-1.2){(b) Color}
  \put(17,-3.2){temperature}
  \put(31.5,-1.2){(c) Jpeg}
  \put(30.3,-3.2){compression}
  \put(43.5,-1.2){(d) Brightness}
  \put(44.8,-3.2){adjustment}
  \put(58.5,-1.2){(e) Contrast}
  \put(59.5,-3.2){adjustment}
  \put(72.0,-1.2){(f) Saturation}
  \put(73.2,-3.2){adjustment}
  \put(85.8,-1.2){(g) Gaussian}
  \put(88.9,-3.2){noise}
  \end{overpic}
  \vspace{2mm}
  \caption{Image distortion pipeline. During training, in order to make our model robust to distortion during transmission, we insert a distortion network that simulates the distortion between the encoder and decoder. We take the encoder output and apply the random transformations shown here before passing the output through the decoder. }
  \label{figdistortion}
\end{figure*}
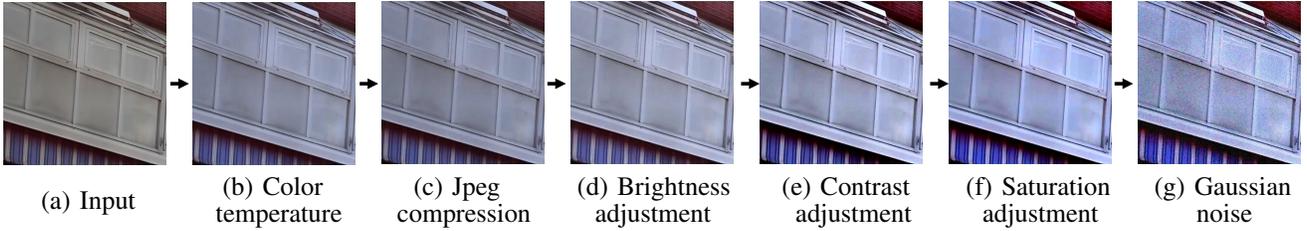

\textbf{Color temperature}: 
% Calculating correlated color temperatures across the entire gamut of daylight and skylight chromaticities
In order to adjust the color temperature of images, we use a Kelvin table which comprises of Kelvin values~\cite{hernandez1999calculating} of different color temperatures.
\begin{equation}
    (r^{'}, g^{'}, b^{'}) = (r \times \frac{r_{t}}{255}, g \times \frac{g_{t}}{255}, b \times \frac{b_{t}}{255}),
\end{equation}
where $r_{t}$, $g_{t}$, $b_{t}$ are the Kelvin values in color temperatures $t$. The color temperature $t$ is randomly selected from:$[6500 - \frac{5500}{\lambda  -\epsilon}, 6500 + \frac{5500}{\lambda  -\epsilon}]$, where $\epsilon$ is the current epoch number, $\lambda$ is the total epoch number of third stage. The meaning behind the article remains the same as stated above.

\textbf{JPEG compression}:
% Jpeg-resistant Adversarial Images
We use the differentiable JPEG compression proposed by~\cite{shin2017jpeg} to approximate rounding function:
\begin{equation}
    rounding(x) =\left\{\begin{matrix}
     x^{3} & |x| < 0.5 \\
     x & |x| \ge 0.5
\end{matrix}\right. .
\end{equation}
The JPEG quality factor is randomly selected from $[60+\frac{40}{\epsilon+1}, 100]$.

\textbf{Brightness, contrast, saturation, and noise}:
For the brightness, contrast and saturation, we use the functions in Kornia. The brightness, contrast and saturation factor is randomly selected from 
$[0-\frac{0.3}{\lambda  - \epsilon}, 0+ \frac{0.3}{ \lambda  - \epsilon}]$, $[0-\frac{0.1}{\lambda  - \epsilon}, 0+\frac{0.1}{\lambda  - \epsilon}]$ and $[0, 0+ \frac{0.1}{\lambda  - \epsilon}]$ respectively. At the same time, we use the Gaussian noise and the standard deviation is randomly selected from $[0,0+ \frac{0.05}{\lambda  - \epsilon}]$.

\subsection{Three-Stage Training and Loss Function}
\label{section:Three}
To strike an optimal balance between decoding accuracy, image quality, and robustness to noise, we have designed a three-stage training process. Each stage focuses on optimizing a specific goal: decoding accuracy, image quality, and robustness to noise, respectively. In the first stage, we prioritize achieving successful watermarking decoding, which comes at the cost of noticeable degradation in both RAW and RGB image quality. Moving to the second stage, we observe improvements in image quality, but this might result in increased vulnerability of the watermarking to distortion. In the third stage, we reinforce the resilience of the watermarking by gradually introducing noise augmentation. It is essential to note that our experimental findings demonstrate that attempting to simultaneously optimize all goals from the beginning leads to non-convergence of the loss function. Hence, our step-by-step approach allows us to effectively enhance different aspects of the watermarking process and achieve the desired trade-offs.

\textbf{The First Stage}:
At this stage, the optimization is solely focused on the RGB image decoder. The loss function for optimizing the decoder is defined as $L_{S1}$:
\begin{equation}
    L_{S1}(M, M_{d}) = L_{dec}(M, M_{d}) = Cross-Entropy(M, M_{d}).
\end{equation}

\textbf{The Second Stage}:
After the initial stage, our model can retrieve the hidden watermarking message from the encoded RGB images generated by the ISP pipeline. However, it should be noted that the appearance of the encoded RGB images closely resembles Quickly Respond (QR) code images. The observed phenomenon can be attributed to the absence of the visual quality constraints imposed on the model. Therefore, in this stage, we use $L_{S2}$ to the second stage as follow: 
\begin{equation}
    \begin{split}
    & L_{S2}(M, M_{d},R,R_{e},I_{o},I_{e}) = \lambda_{1} L_{dec}(M, M_{d}) + \lambda_{2}L_{2}(R,R_{e})  \\
    & + \lambda_{3}L_{2}(I_{o},I_{e}) + \lambda_{4}L_{P}(I_{o},I_{e}) + \lambda_{5}L_{d}(I_{o},I_{e}),
    \end{split}
\end{equation}
where $L_{2}$ is L2 norm loss, $L_{P}$ is the LPIPS~\cite{zhang2018perceptual} loss and $L_{d}$ is the discriminator loss.
$\lambda_{1}$, $\lambda_{2}$, $\lambda_{3}$, $\lambda_{4}$ and $\lambda_{5}$ are hyperparameter set to 2, 1, 1, 1, and 1, respectively.

\textbf{The Third Stage}:
Following the aforementioned two stages of training, the model has acquired the ability to encode the information into RAW images and extract the information from RGB images, simultaneously preserving the visual fidelity of both RAW and RGB images. However, the current watermarking is not robust to different distortions (\textit{e.g.,} Gaussian Noise, JPEG compression). 
To effectively extract the embeded watermarking from the distorted RGB images. We train our encoder and decoder using $L_{S3}$ as follows:
\begin{equation}
    \begin{split}
    & L_{S3}(M, M^{'}_{d},R,R_{e},I_{o},I_{e}) = \lambda_{1} L_{dec}(M, M^{'}_{d}) + \lambda_{2}L_{2}(R,R_{e})  \\
    & + \lambda_{3}L_{2}(I_{o},I_{e}) + \lambda_{4}L_{P}(I_{o},I_{e}) + \lambda_{5}L_{d}(I_{o},I_{e}).
    \end{split}
\end{equation}
The $L_{S3}$ is very similar to the $L_{S2}$, only the second parameter is different, where $M^{'}_{d}$ means decoded binary message from distorted RGB images and the hyperparameters setting is same as second stage. 
\begin{equation}
    M^{'}_{d} = D_{\theta_{D}}(\mathbf{N}(I_{e})),
\end{equation}
where $\mathbf{N}$ is present the distortion network, which can help the encoder to add anti-distortion watermarking into RAW images.

\begin{figure*}[h]
  \centering
  \renewcommand{\tabcolsep}{1.0pt} %
  \renewcommand{\arraystretch}{0.5}
    \begin{tabular}{ccc}
        \includegraphics[width=0.333\textwidth]{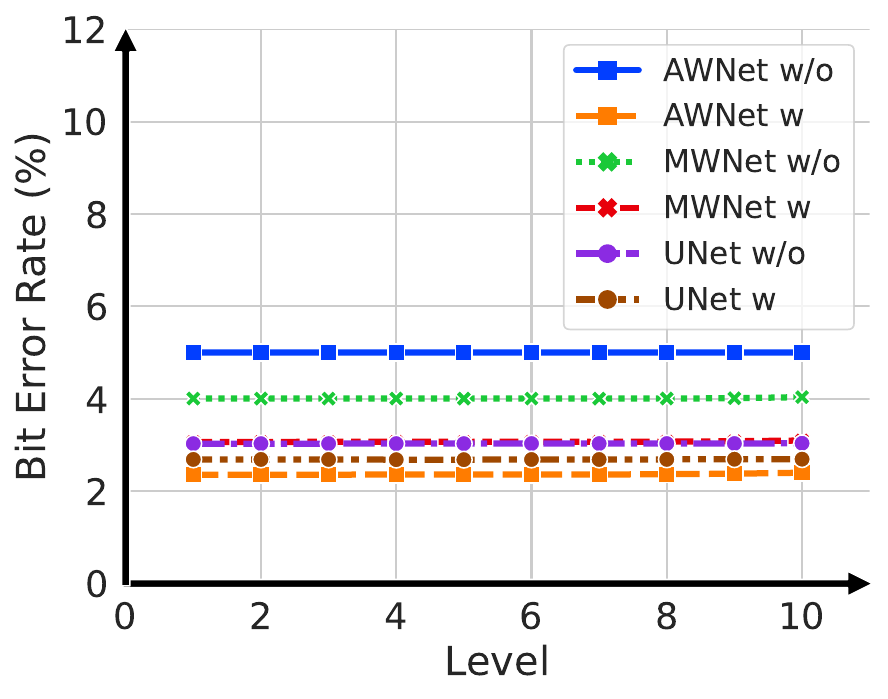} & 
        \includegraphics[width=0.333\textwidth]{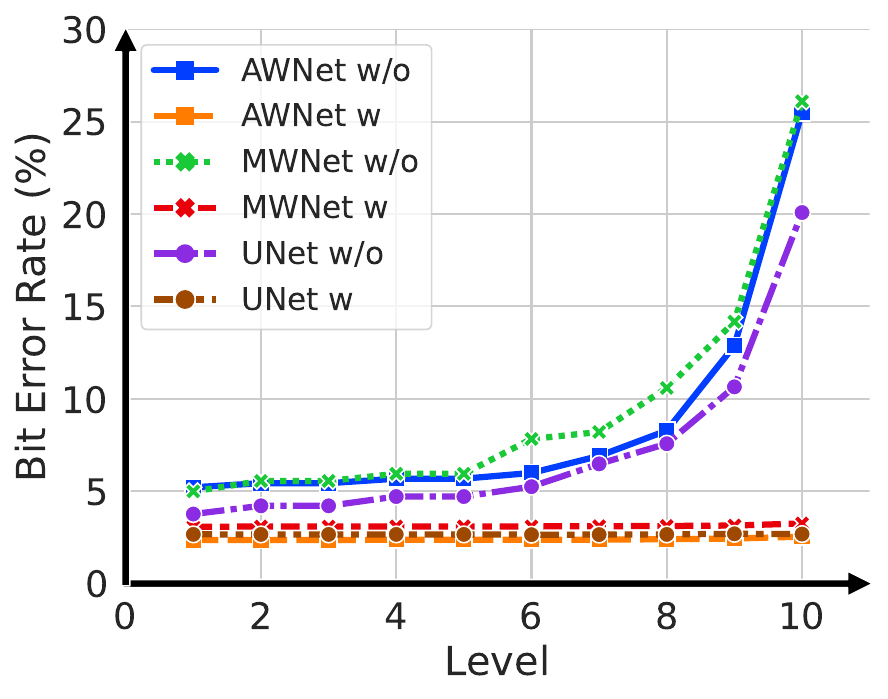} & 
        \includegraphics[width=0.333\textwidth]{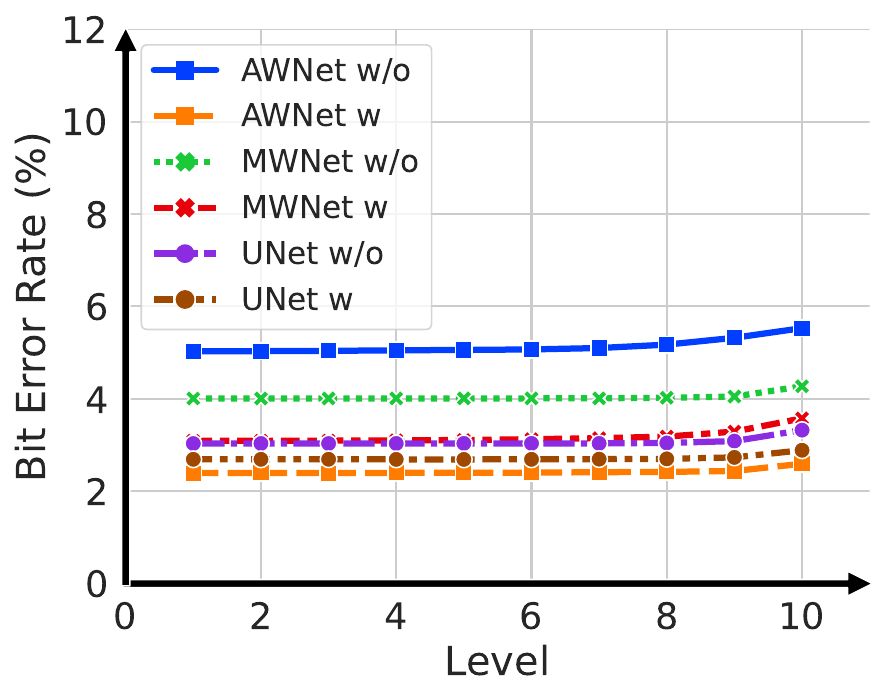} \\
        (a) Color Temperature & (b) Jpeg Compression & (c) Brightness Adjustment\\
        \includegraphics[width=0.333\textwidth]{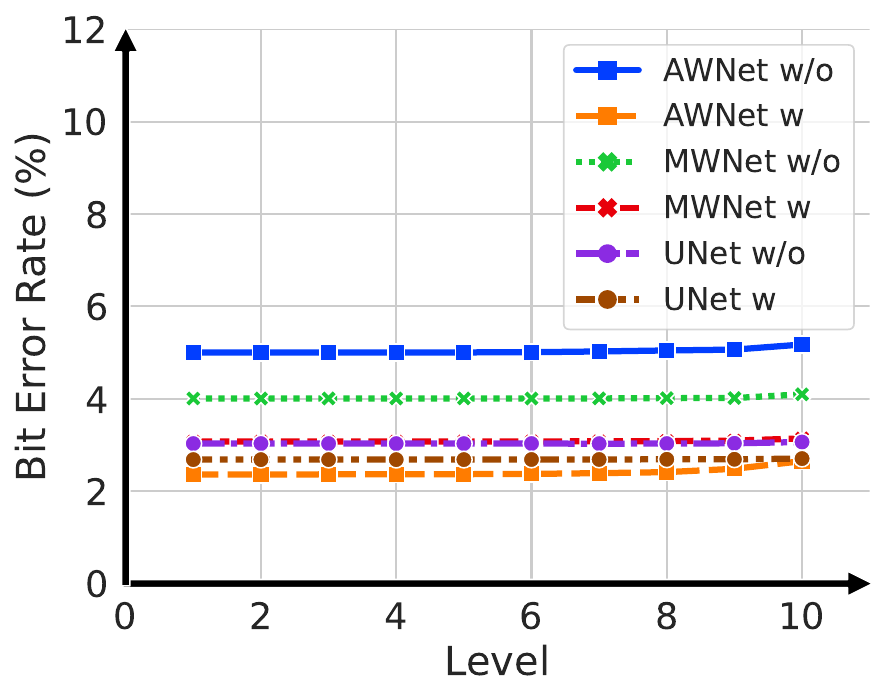} &
        \includegraphics[width=0.333\textwidth]{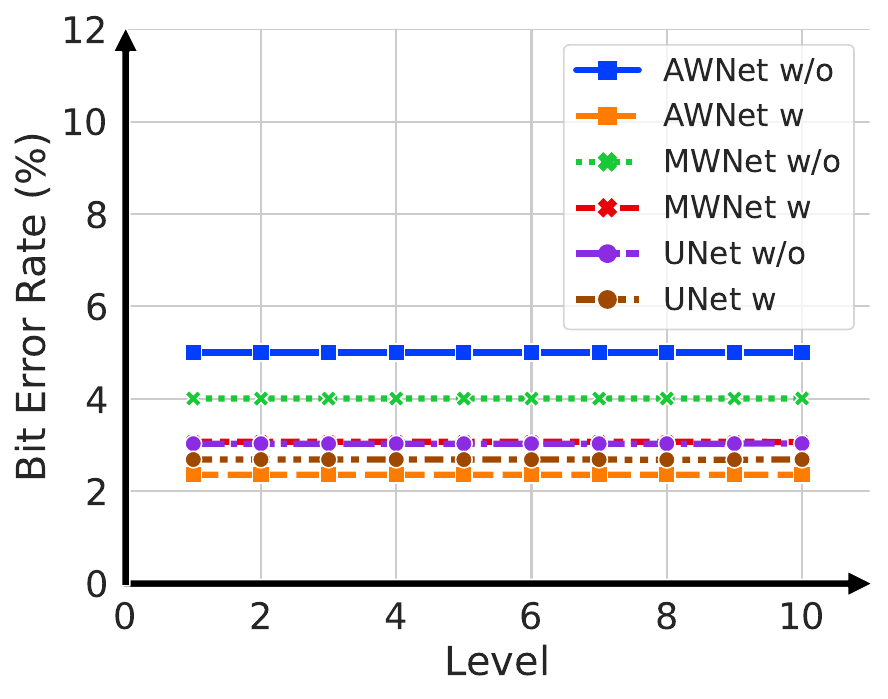} &
        \includegraphics[width=0.333\textwidth]{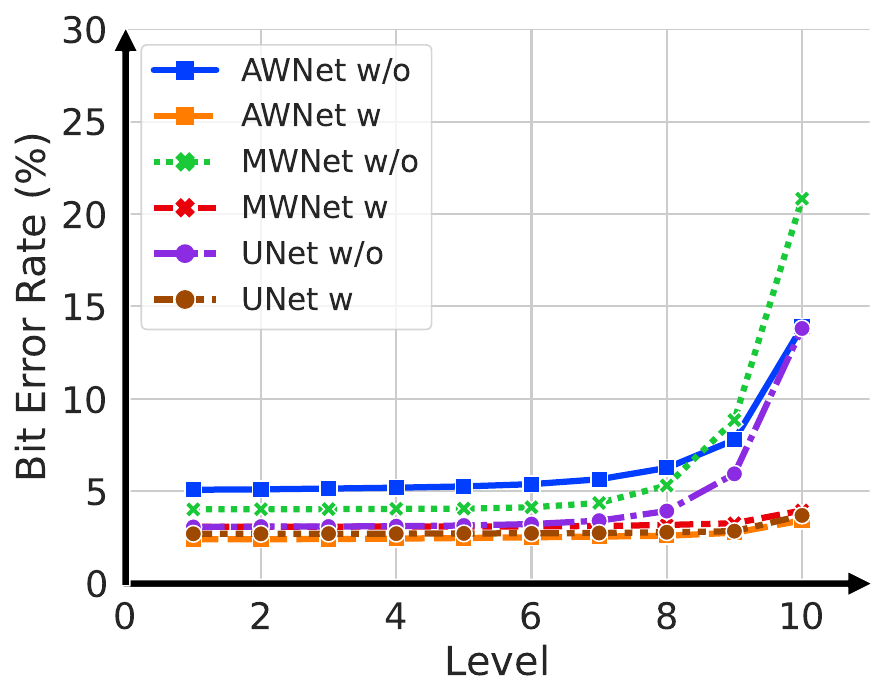} \\ 
        (d) Contrast Adjustment & (e) Saturation Adjustment & (f) Gaussian Noise \\
    \end{tabular}
  % \caption{Bit Error Rate (BER) of the decoding results of different models under diverse levels of various distortions. For the model, First key: A: AWNet, M: MW-ISPNet, U: UNet; Second key: C: Combined encoder; Third key: W: training with distortion network, O: training without distortion network.}
  \caption{Bit Error Rate (BER) of the decoding results of different models (The combined encoder is used in the training of the above models.) under diverse levels of various distortions. MW-ISPNET is abbreviated as MWNet.}
  \label{figroubusttest}
\end{figure*}

%% file: Contents/Experiments.tex
\section{Experiments}
\label{sec:Experiments}
Within this section, we give the details about out experimental settings and conduct extensive experiments.

\subsection{Experimental Setting}

\textbf{Dataset}:
In our experiments, we primarily utilized two datasets: the ZRR dataset~\cite{ignatov2020replacing} and the SR-RAW dataset~\cite{zhang2019zoom}. The ZRR dataset served as the training dataset, comprising 47,863 paired images. This dataset consists of RAW images of size $448 \times 448 \times 1$, captured by Huawei p20, and RGB images of size $448 \times 448 \times 3$, captured and processed by the Canon 5D Mark IV camera. The test set of the ZRR dataset was used to conduct various experiments, including ablation and robustness testing experiments.

However, since the RAW images in the aforementioned dataset are stored in PNG format, they lack the camera parameters saved during shooting. Consequently, traditional ISP pipelines cannot directly convert them to RGB format. To overcome this limitation, we conducted traditional ISP test experiments using the test set of the SR-RAW dataset. The test set collected 50 scenes of images, with the authors of dataset taking 7 photos at different focal lengths (24, 35, 50, 70, 100, 150, and 240 mm) in each scene. We randomly selected 100 of these images for the traditional ISP test experiment.

\textbf{Hyper-Parameter Setting and Implementation Details}:
As highlighted in Section \ref{section:Three}, the training phase of our model is segregated into three stages. Specifically, we conducted 5,5 and 10 epochs for the first stage, second stage, and the third stage respectively. 

Two optimizers were used in the model training process. The Adam optimizer was assigned to the task of optimizing the RAW image encoder and the RGB image decoder. The RMSProp was to optimize the discriminator. The learning rate of both optimizers is $5 \times 10^{-5}$. The batch size is set to 4.

For our watermarking, we elected to generate the 100-bit binary strings randomly during the training phase. In the testing phase, we restricted the embeded watermarking to 56 bits, subsequently incorporating error-correcting codes using Bose–Chaudhuri–Hocquenghem (BCH) codes~\cite{bose1960class} to fill the remaining 44 bits, ultimately arriving at a 100-bit watermarking.

It is important to note that we have made certain adjustments to the number of submodules proposed by MW-ISPNet and AWNet in comparison to their respective public models. This reduction in the number of submodules has proven to be beneficial for the convergence of the entire framework, resulting in faster training. We used publicly available code for training purposes but made specific modifications to the models, such as adjusting the number of modules and channel features. Specifically, for AWNet, we set the number of GCRDB submodules to [1, 1, 1, 2, 3]. As for MWNet, we configured the feature channels to [32, 64, 64]. On the other hand, the UNet ISP employs a similar framework to the encoder, and the training process closely resembles that of AWNet.

\begin{table}
\renewcommand{\arraystretch}{1.1}
  \caption{Traditional ISP pipeline test results. For the ISP (traditional ISP pipelines), key: R-A: auto white balance; R-C: camera white balance; R-D: daylight white balance. For the ISPT (deep ISP pipelines in training of models). For the distortion, $\checkmark$ indicating the distortion network is used in training of model and $\times$ indicating not. }
  \label{ISP}
\begin{tabular}{c|c|c|cccc}
\toprule
\multicolumn{1}{c|}{ISP} & ISPT & Distortion & PSNR   & SSIM  & BER    & SER     \\ \midrule
\multirow{6}{*}{R-A}    & AWNet            & $\times     $       & 32.653 & 0.811 & 3.73\% & 8.33\%  \\ 
                        & AWNet           & \checkmark          & 30.794 & 0.857 & 2.81\% & 0.02\%  \\
                        & MWNet       & $\times     $        & 27.047 & 0.689 & 5.63\% & 0.27\%  \\
                        & MWNet       & \checkmark           & 27.884 & 0.691 & 3.10\% & 0.00\%  \\
                        & UNet            & $\times     $        & 31.736 & 0.891 & 3.85\% & 14.58\% \\
                        & UNet            & \checkmark           & 31.477 & 0.904 & 2.50\% & 0.00\%  \\ \hline
\multirow{6}{*}{R-C}    & AWNet           & $\times     $        & 32.704 & 0.812 & 3.67\% & 10.42\% \\ 
                        & AWNet           & \checkmark           & 30.866 & 0.859 & 3.08\% & 4.17\%  \\
                        & MWNet       & $\times     $        & 27.110 & 0.698 & 5.60\% & 0.25\%  \\
                        & MWNet       & \checkmark           & 28.060 & 0.693 & 2.77\% & 0.00\%  \\
                        & UNet            & $\times     $        & 31.474 & 0.905 & 3.71\% & 12.50\% \\
                        & UNet            & \checkmark           & 31.870 & 0.892 & 2.42\% & 0.00\%  \\ \hline
\multirow{6}{*}{R-D}    & AWNet           & $\times     $        & 32.939 & 0.820 & 3.40\% & 4.17\%  \\ 
                        & AWNet           & \checkmark          & 30.954 & 0.860 & 2.92\% & 2.08\%  \\
                        & MWNet       &$\times     $        & 27.222 & 0.679 & 5.42\% & 16.67\% \\
                        & MWNet       & \checkmark           & 28.348 & 0.707 & 2.77\% & 4.17\%  \\
                        & UNet            & $\times     $        & 31.727 & 0.907 & 2.56\% & 0.00\%  \\
                        & UNet            & \checkmark           & 31.947 & 0.892 & 4.25\% & 0.15\%  \\
    \bottomrule
  \end{tabular}
\end{table}

\textbf{Evaluation Metrics}:
In the subsequent experiments, we assessed the efficacy of our model in terms of its concealment and robustness capabilities. In regards to concealment, three metrics were employed, namely Peak Signal to Noise Ratio (PSNR), Structural Similarity (SSIM)~\cite{1284395}, and Learned Perceptual Image Patch Similarity (LPIPS)~\cite{zhang2018perceptual}. It is noteworthy that, in this paper, version 0.1 of LPIPS trained on the AlexNet network was utilized. 

On the other hand, the evaluation of robustness involved the use of two metrics: Bit Error Rate (BER), and String Error Rate (SER):
\begin{equation}
    BER = \frac{E_{bit}}{A_{bit}}, SER = \frac{E_{str}}{A_{str}},
\end{equation}
in which $E_{bit}$ represents the number of bits in error, and $A_{bit}$ represents the total number of bits to be hidden. The meaning of $E_{str}$ and $A_{str}$ is similar. It is noteworthy that, in our experiment, the string was subjected to BCH coding. Consequently, even if the number of erroneous bits is small, the string can still be accurately restored.

\subsection{Robustness Test}
In this section, we test the robustness of our model to different distortions. We test the distortions used in training,  For each distortion category, we set 10 different distortion levels (The larger the number, the greater the distortion). 
1) Color temperature adjustment: color temperature is uniform sampling in $ T \sim U[6500 - \frac{5500}{10-\varepsilon}, 6500 + \frac{5500}{10-\varepsilon}] $;
2) JPEG Compression: JPEG quality is uniform sampling in $J \sim  U[60 + \frac{40}{\varepsilon+1}, 100]$;
3) Brightness and contrast adjustment: affine histogram rescaling $mx + b$ with  $ m \sim U[0-\frac{0.1}{10-\varepsilon}, 0+\frac{0.1}{10-\varepsilon}]$ and $b \sim U[0-\frac{0.3}{10-\varepsilon}, 0+ \frac{0.3}{10-\varepsilon}]$;
4) Noise and Saturation adjustment: we use a Gaussian noise (sampling the standard deviation $\sigma \sim U[0,0+ \frac{0.05}{10-\varepsilon}]$) and saturation factor is uniform sampling in $S \sim  U[0, 0+ \frac{0.1}{10-\varepsilon}]$.
where $\varepsilon $ is the distortion level.
The specific results of the experiment are shown in Fig \ref{figroubusttest}.

\subsection{Traditional ISP Pipeline Test}
% #SR-RAW
To assess the robustness of the proposed method against the traditional ISP pipeline, a validation experiment was conducted. To achieve the necessary processing from RAW to RGB images, In the experiment, the traditional ISP pipeline we selected is Rawpy library. The Rawpy library, built on the foundation of LibRaw, serves as the underlying implementation for our work. This library offers a range of image processing operations that enable the conversion of RAW images into RGB images. These operations include essential tasks like white balance correction, color space conversion, and brightness adjustment. Given the capabilities and functionality of rawpy, we consider it to effectively represent traditional ISP methodologies. The {R}awpy library has three white balance models: R-A (it \textbf{A}utomatically calculates the white balance), R-C (it uses the as-shot white balance values of \textbf{C}amera), and R-D (it uses \textbf{D}aylight white balance correction). The encoding and decoding processes of the proposed method remains unchanged throughout the experiment.
One hundred unique images were randomly selected from the SR-RAW dataset, with the condition that these images were disjoint from the training set used in the proposed method. The random 56-bit messages, after being subjected to using BCH coding to 100-bit messages, were embedded within each of these selected images. The results of this experiment are presented in Table \ref{ISP}. A total of six models were subjected to testing, which are all using combined encoder. 
Examples of watermarked images generated by different models in this experiment are shown in Fig \ref{figisp}.
The experimental results demonstrate the robustness of proposed RAWIW framework against the traditional ISP pipelines, and using the distortion network in training can improve the decoding accuracy of the watermarking for the traditional ISP pipeline.

\begin{table*}
    \renewcommand{\arraystretch}{1.05}
    \renewcommand{\tabcolsep}{10.5pt} %
    \caption{Ablation study results with distortion network. `R', `D' and `C' stand for RAW encoder, demosaicing encoder and combined encoder respectively. Best in \textcolor{red}{red} and second in \textcolor{blue}{blue}.}
    \label{Table2}
    \begin{tabular}{c|c|ccc|cc|cc|c|c}
    \toprule
    % \multicolumn{1}{l}{ISP}  & Encoder  & $PSNR_{raw}\uparrow$ & $PSNR_{gt}\uparrow$ & $PSNR_{isp}\uparrow$ & $LPIPS_{gt} \downarrow$ & $LPIPS_{isp}\downarrow$ & $SSIM_{gt}\uparrow$ & $SSIM_{isp}\uparrow$ & $BER\downarrow$ & $SER\downarrow$ \\ \midrule
\multirow{2}{*}{ISP} & \multirow{2}{*}{Encoder} & \multicolumn{3}{c|}{PSNR$\uparrow$} & \multicolumn{2}{c|}{LPIPS$\downarrow$}  & \multicolumn{2}{c|}{SSIM$\uparrow$}  & \multirow{2}{*}{BER$\downarrow$} & \multirow{2}{*}{SER$\downarrow$} \\ \cline{3-9}
   &   & RAW    & GT   & ISP     & GT     & ISP    & GT     & ISP      & &  \\\midrule
    \multirow{3}{*}{MW-ISPNet~\cite{ignatov2020aim}} & R    & 45.389 & 20.164 &  28.979 &  0.154 & 0.068 & 0.804 & 0.975 & \textcolor{blue}{1.035\%} &  \textcolor{blue}{0.083\%}            \\ 
                               & D    & 37.060 &  20.487 & 31.078 &  0.134 &  0.051 & 0.804 &  0.974 & 3.006\% &  \textcolor{blue}{0.083\%} \\ 
                               & C    & 37.890 & 20.144 & 29.169 &  \textcolor{red}{0.113} &  \textcolor{red}{0.020} & 0.804 &  0.981 & 1.155\% & 1.080\% \\ \midrule

    \multirow{3}{*}{Unet~\cite{ronneberger2015u}}      & R    & \textcolor{red}{47.581}  & 19.088 & 32.242 &  0.174 &  0.040 & 0.792 & \textcolor{blue}{0.988} & 1.088\% &  0.332\% \\  
                               & D    & 44.858 & 18.486 & 30.749 &  0.180 &  0.038 & 0.787 & 0.986 & 4.002\% &  0.332\% \\ 
                               & C    & 44.859  &  19.637 & 31.693 & 0.154  & 0.036  & 0.797  &  0.987  & 2.029\%   & 0.166\%     \\ \midrule
    \multirow{3}{*}{AWNet~\cite{dai2020awnet}}     & R   & 44.347 &  20.507 &  31.407 & \textcolor{blue}{0.116}  & \textcolor{blue}{0.024} & 0.812  & 0.978 & 2.014\%  & 0.166\% \\
                               & D    & \textcolor{blue}{46.399} & \textcolor{blue}{20.782}  &  \textcolor{blue}{32.751} &  0.143 & 0.056 &  \textcolor{blue}{0.816} &  0.984 & 2.086\% & 0.498\% \\ 
                               & C    & 40.277 &  \textcolor{red}{21.046} & \textcolor{red}{35.143}  & 0.144 & 0.057 & \textcolor{red}{0.819} &  \textcolor{red}{0.989} & \textcolor{red}{0.880\%} &  \textcolor{red}{0.000\%} \\ 
                               \bottomrule
    \end{tabular}
\end{table*}

\begin{table*}
    \renewcommand{\arraystretch}{1.05}
    \renewcommand{\tabcolsep}{10.5pt} %
    \caption{Ablation study results without distortion network. `R', `D' and `C' stand for RAW encoder, demosaicing encoder and combined encoder respectively. Best in \textcolor{red}{red} and second in \textcolor{blue}{blue}.}
    \label{Table3}
    \begin{tabular}{c|c|ccc|cc|cc|c|c}
    \toprule
    % \multicolumn{1}{l}{ISP}  & Encoder  & $PSNR_{raw}\uparrow$ & $PSNR_{gt}\uparrow$ & $PSNR_{isp}\uparrow$ & $LPIPS_{gt} \downarrow$ & $LPIPS_{isp}\downarrow$ & $SSIM_{gt}\uparrow$ & $SSIM_{isp}\uparrow$ & $BER\downarrow$ & $SER\downarrow$ \\  \midrule
\multirow{2}{*}{ISP} & \multirow{2}{*}{Encoder} & \multicolumn{3}{c|}{PSNR$\uparrow$} & \multicolumn{2}{c|}{LPIPS$\downarrow$}  & \multicolumn{2}{c|}{SSIM$\uparrow$}  & \multirow{2}{*}{BER$\downarrow$} & \multirow{2}{*}{SER$\downarrow$} \\ \cline{3-9}
   &   & RAW    & GT   & ISP     & GT     & ISP    & GT     & ISP      & &  \\\midrule
    \multirow{3}{*}{MW-ISPNet~\cite{ignatov2020aim}} & R         & 45.240 & 20.737 & 36.762 & 0.111 & 0.012 & 0.814 & 0.983 & 4.071\% & 1.495\% \\
                               & D         & 43.531 & 20.728 & 39.480 & 0.109 & 0.008 & 0.814 & 0.984 & 2.354\% & 2.658\% \\ 
                               & C         & 47.623 & 20.871 & 45.018 & \textcolor{red}{0.106} & \textcolor{red}{0.001} & \textcolor{blue}{0.816} & \textcolor{red}{0.989} & \textcolor{red}{2.179\%} & 0.831\% \\  
                               \midrule
    \multirow{3}{*}{Unet~\cite{ronneberger2015u}}      & R         & \textcolor{blue}{50.110} & 19.289 & 38.834 & 0.163 & 0.006 & 0.798 & \textcolor{blue}{0.988} & 3.034\% & 0.664\% \\  
                               & D         & 48.806 & 19.449 & 40.052 & 0.156 & 0.006 & 0.799 & 0.987 & \textcolor{blue}{2.252\%} & 1.495\%         \\ 
                               & C         & 44.292 & 19.402 & 40.814 & 0.160 & 0.004 & 0.798 & \textcolor{blue}{0.988} & 3.686\% & \textcolor{red}{0.166\%}         \\ \midrule
    \multirow{3}{*}{AWNet~\cite{dai2020awnet}}     & R         & \textcolor{red}{50.966} & \textcolor{red}{21.316} & 46.145 & \textcolor{blue}{0.108} & \textcolor{blue}{0.002} & \textcolor{red}{0.826} & \textcolor{red}{0.989} & 3.100\% & 0.748\% \\
                               & D         & 49.489 & 21.281 & \textcolor{blue}{46.294} & 0.109 & 0.003 & \textcolor{red}{0.826} & \textcolor{blue}{0.988} & 4.041\% & \textcolor{blue}{0.498\%} \\        
                               & C         & 46.181 & \textcolor{blue}{21.311} & \textcolor{red}{47.303} & \textcolor{blue}{0.108} & \textcolor{blue}{0.002} & \textcolor{red}{0.826} & \textcolor{blue}{0.988} & 3.359\% & 2.409\% \\       
                               \bottomrule
    \end{tabular}
\end{table*}

\begin{table}[]
\renewcommand{\arraystretch}{1.1}
\renewcommand{\tabcolsep}{10.5pt} %
\caption{Comparison results with state-of-the-art RGB image watermarking methods. Best in \textcolor{red}{red} and second in \textcolor{blue}{blue}.}
\label{Table4}
\begin{tabular}{c|cc|cc}
\toprule
model & PSNR $\uparrow$  & SSIM $\uparrow$  & BER $\downarrow$    & SER $\downarrow$    \\
\midrule
Stegastamp\cite{tancik2020stegastamp} & 29.31  & 0.927 & \textcolor{blue}{0.230\%} & \textcolor{red}{0.000\%} \\\midrule
RIHOOP\cite{jia2020rihoop}    & 29.42  & 0.940 & \textcolor{red}{0.120\%} &\textcolor{red}{ 0.000\%} \\\midrule
RAWIW(M)                  & 29.169 & 0.981 & 1.155\% & 1.080\% \\\midrule
RAWIW(U)                 & \textcolor{blue}{31.693} & \textcolor{blue}{0.987} & 2.029\% & \textcolor{blue}{0.166\%} \\\midrule
RAWIW(A)                 & \textcolor{red}{35.143} & \textcolor{red}{0.989} & 0.880\% & \textcolor{red}{0.000\%} \\
\bottomrule
\end{tabular}

\end{table}

\subsection{Ablation Study}
\label{section:abst}
In this section, we conduct ablation experiments to compare watermarking concealment and robustness metrics under different ISP pipeline, encoders and  distortions. We use the test data in the ZRR dataset for testing, a total of 1204 paired images. Throughout the ablation experiment, we trained a total of 18 ($3 \times 3 \times 2$) models, using three ISP networks, UNet, MW-ISPNet, and AWNet, three encoder structures, and whether to use a distortion network.
Our model can be trained to store different numbers of bits. We settle on a message length of 100 bits in our experiment as it provides a good compromise between image quality and information transfer. The watermarking is random 100 bits binary string in training and 100 bits binary string encoded by random 56 bits binary through BCH coding. We can recover watermarking in the case of a few bit errors by using BCH coding.
The experimental results in Table ~\ref{Table2} and ~\ref{Table3} show that: 
1) The ISP network adopts the AWNet structure to make the watermarking concealment better.
2) The combined encoder structure is better than the RAW structure and demosaicing structure.
3) The distortion network will reduce the concealment and improve the robustness of watermarking. Distortion networks enable encoders to add more ``obvious'' watermarking to RAW images against distortion caused by distortion networks. 

\subsection{Comparison with SOTA}

In this section, we have conducted a comparative experiment between two state-of-the-art RGB image watermarking methods: Stegastamp and RIHOOP. To ensure consistency and accurate evaluation, the experiment settings were derived from the ablation experiment \ref{section:abst}, and the RGB image watermarking methods encoded watermarking into RGB images generated by ISP. The corresponding results are presented in Table \ref{Table4}. It is evident that both our RAW image watermarking and RGB watermarking methods can maintain good visual quality while preserving high decoding accuracy. However, it is worth noting that our RAW image watermarking uses RAW images as covers, which have one-third the capacity of RGB images. As a result, the information hiding density of our method is higher compared to the RGB watermarking methods.

\begin{figure*}[!htbp]
  \centering
  \renewcommand{\tabcolsep}{1.0pt} %
    \begin{tabular}{ccccccccc}
        \multirow{2}{*}[4.3em]{\includegraphics[width=0.21\linewidth]{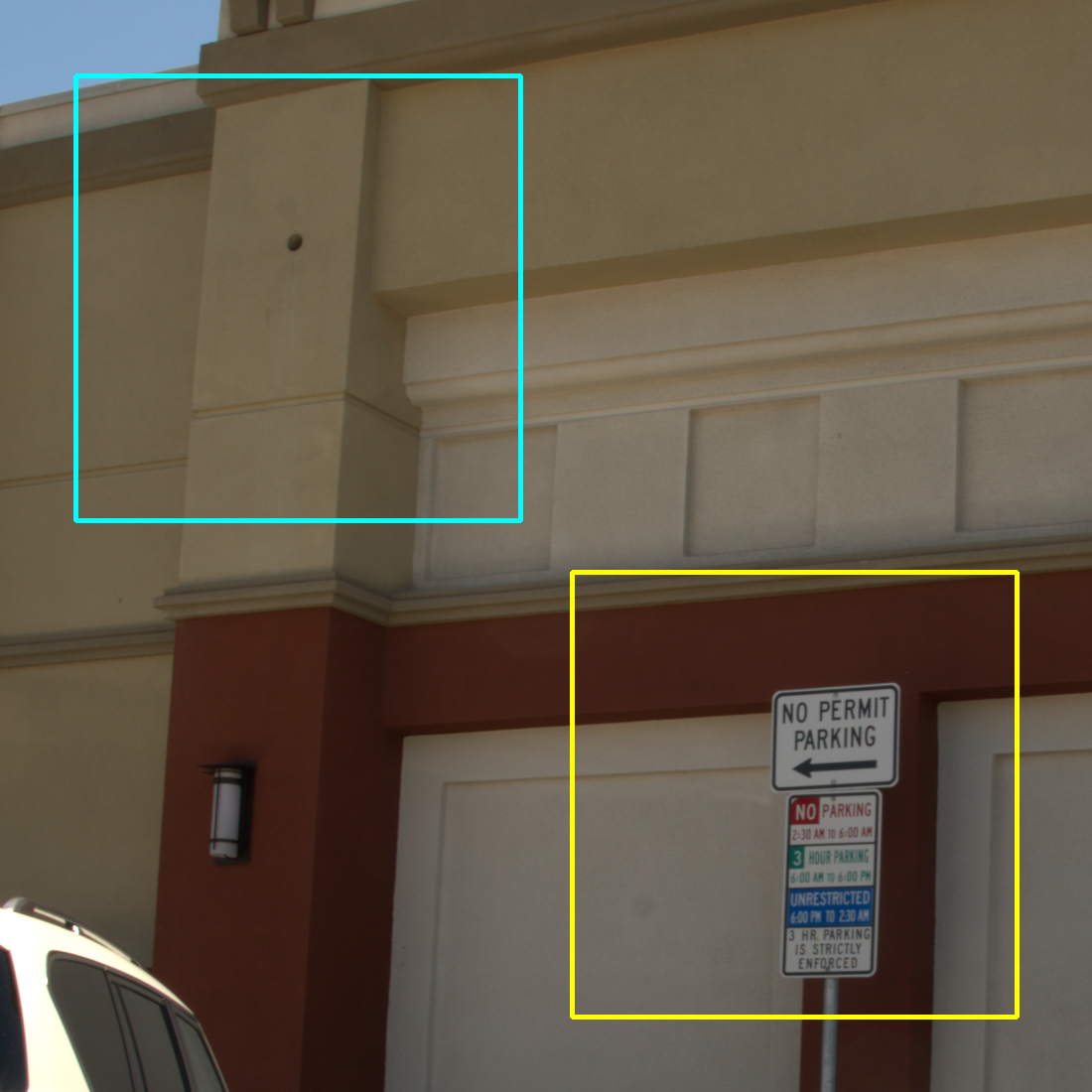}} & 
        \includegraphics[width=0.1\linewidth]{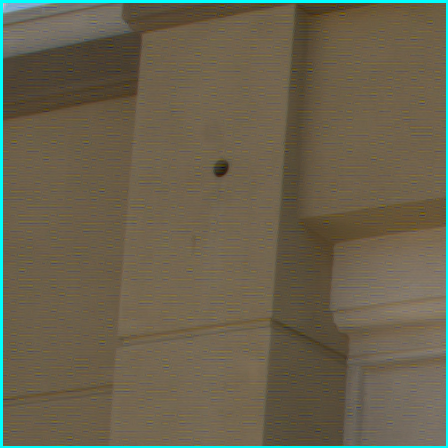} &
        \includegraphics[width=0.1\linewidth]{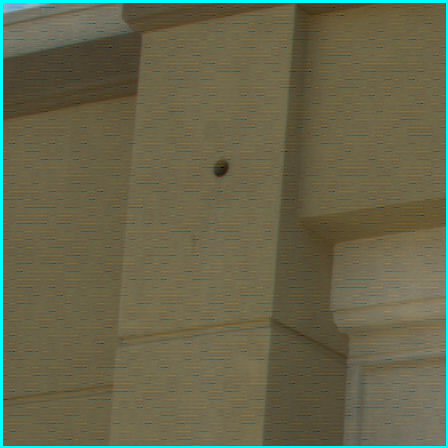} &
        \includegraphics[width=0.1\linewidth]{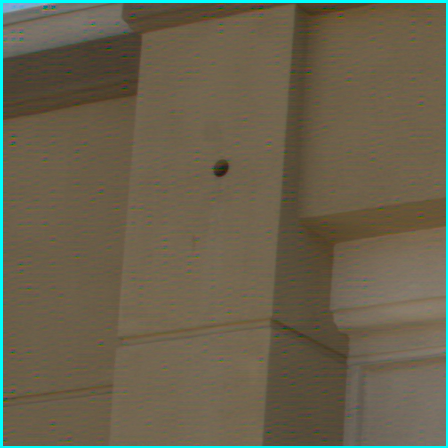} &
        \includegraphics[width=0.1\linewidth]{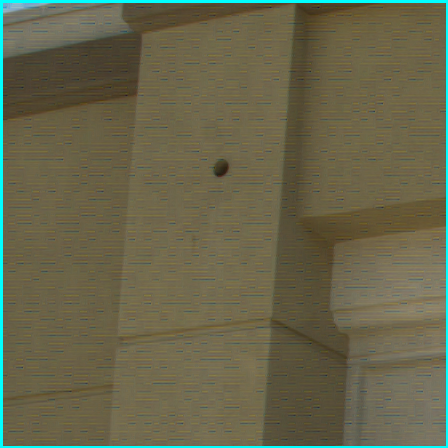} &
        \includegraphics[width=0.1\linewidth]{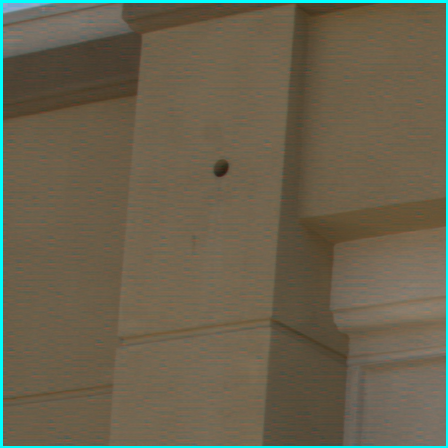} &
        \includegraphics[width=0.1\linewidth]{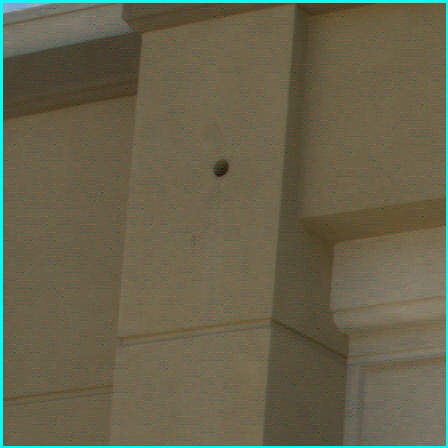} &
        \includegraphics[width=0.1\linewidth]{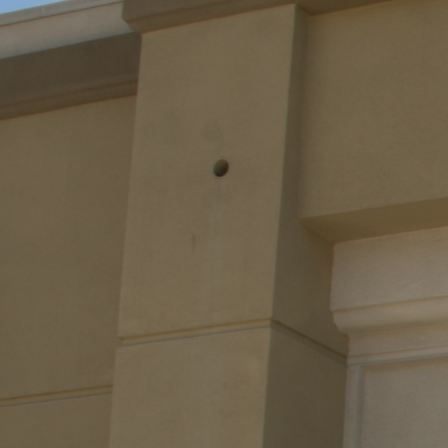} \\ 

        & 
        \includegraphics[width=0.1\linewidth]{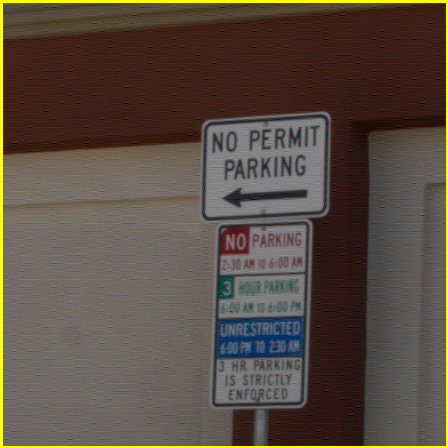} &
        \includegraphics[width=0.1\linewidth]{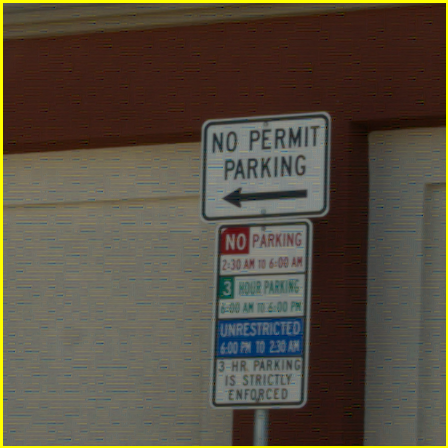} &
        \includegraphics[width=0.1\linewidth]{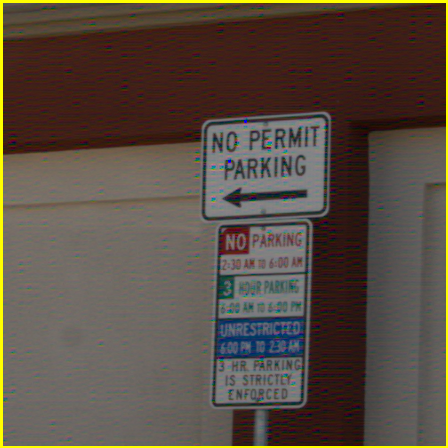} &
        \includegraphics[width=0.1\linewidth]{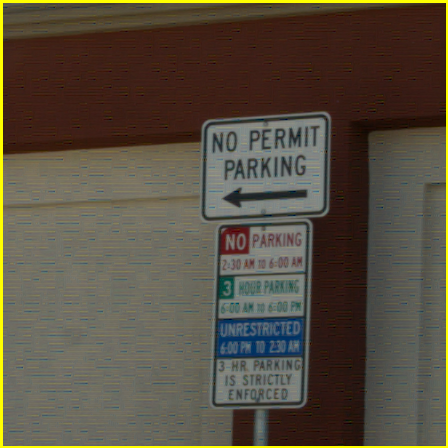} &
        \includegraphics[width=0.1\linewidth]{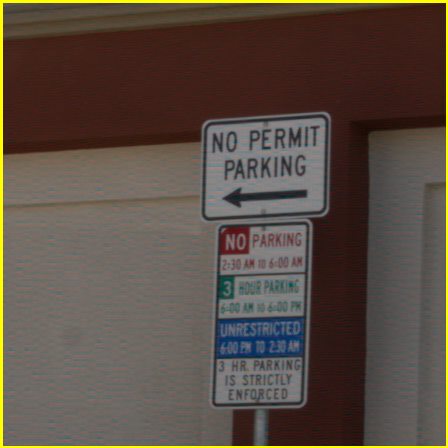} &
        \includegraphics[width=0.1\linewidth]{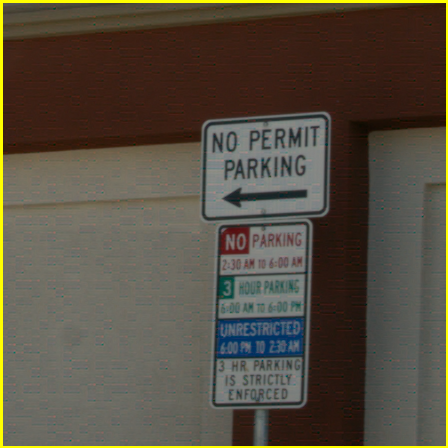} &
        \includegraphics[width=0.1\linewidth]{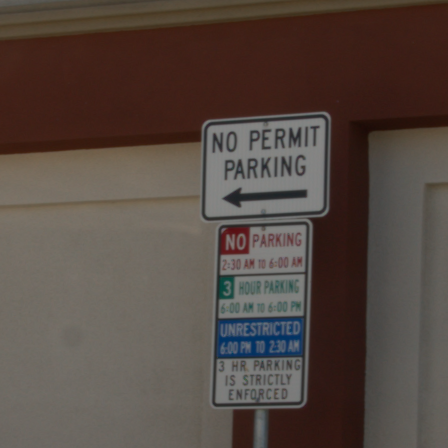} \\ 
        
        \multirow{2}{*}[4.3em]{\includegraphics[width=0.21\linewidth]{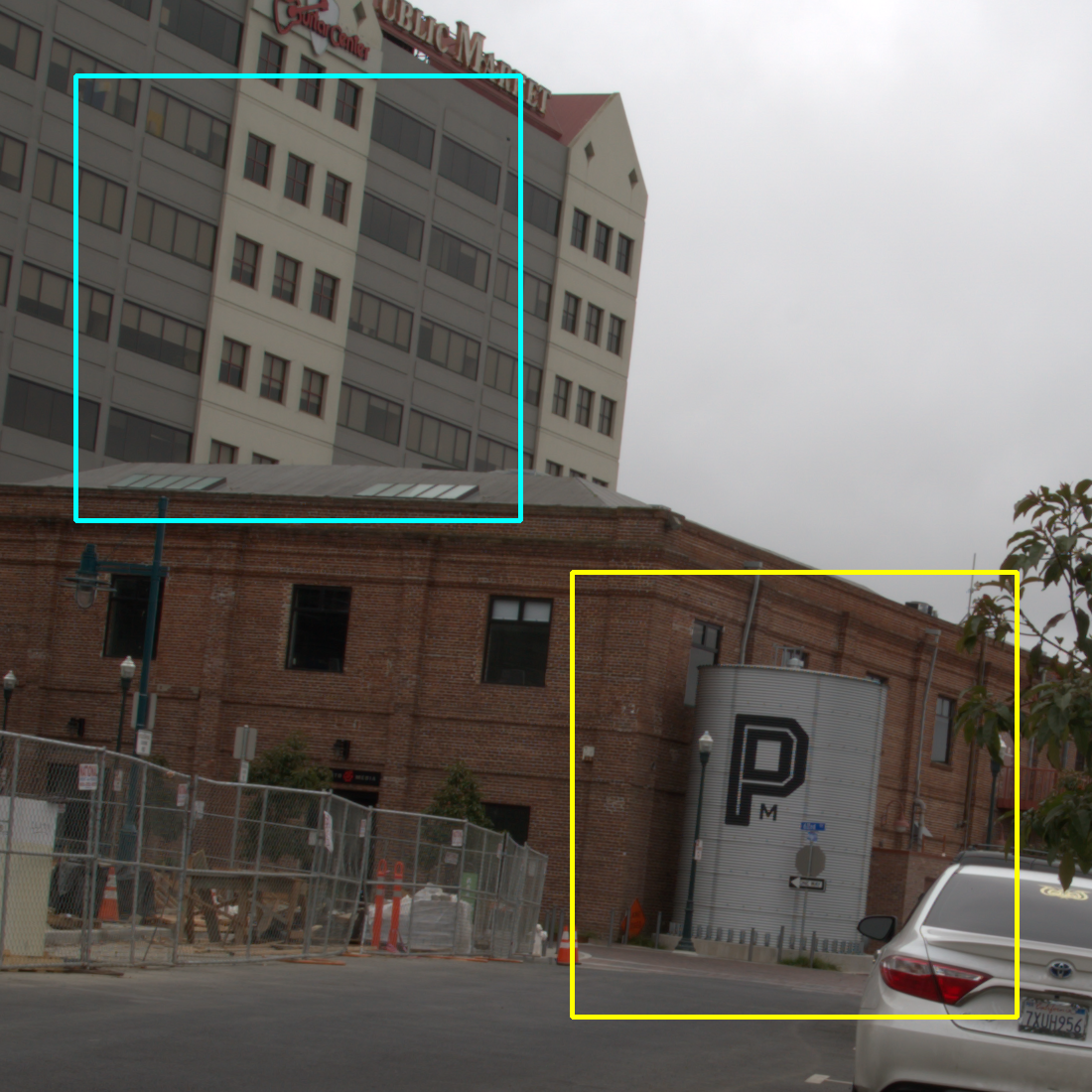}} & 
        \includegraphics[width=0.1\linewidth]{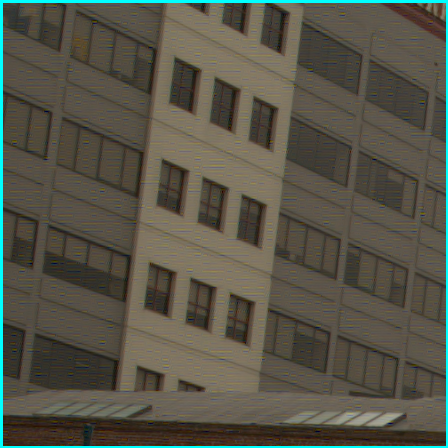} &
        \includegraphics[width=0.1\linewidth]{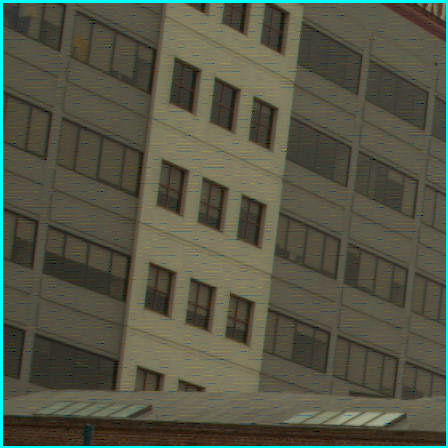} &
        \includegraphics[width=0.1\linewidth]{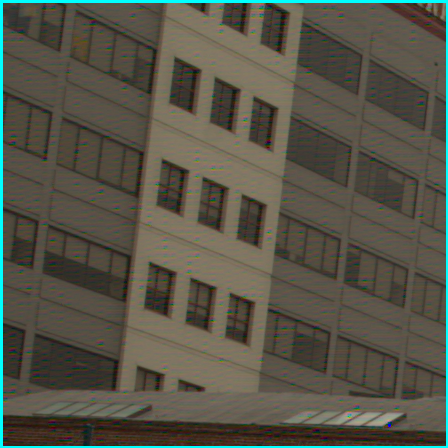} &
        \includegraphics[width=0.1\linewidth]{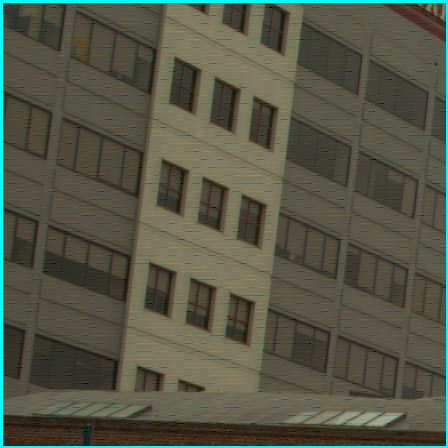} &
        \includegraphics[width=0.1\linewidth]{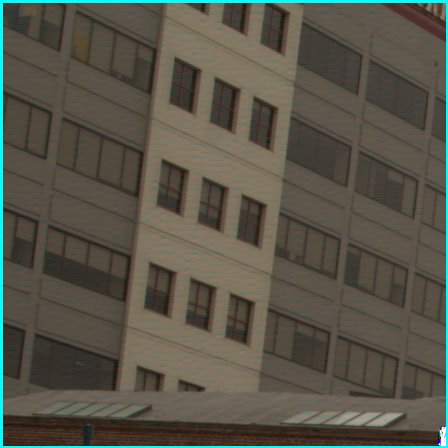} &
        \includegraphics[width=0.1\linewidth]{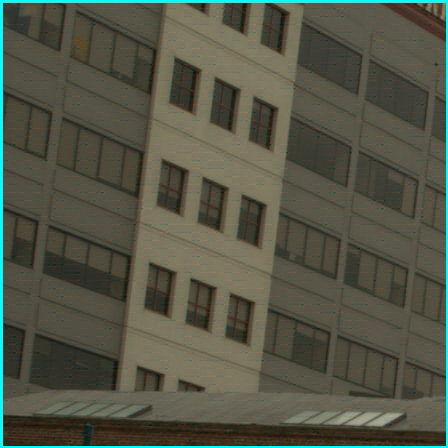} &
        \includegraphics[width=0.1\linewidth]{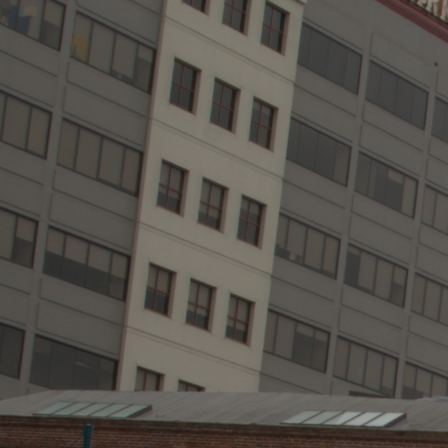} \\ 

        & 
        \includegraphics[width=0.1\linewidth]{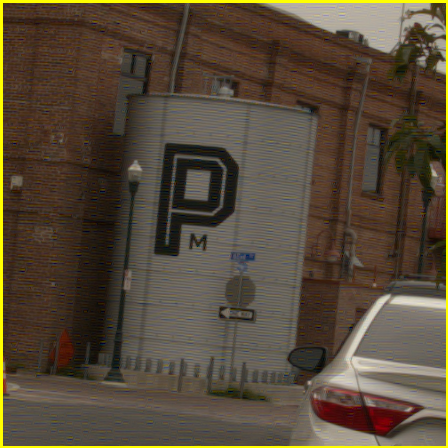} &
        \includegraphics[width=0.1\linewidth]{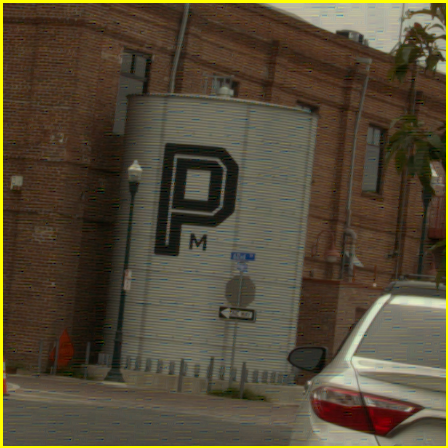} &
        \includegraphics[width=0.1\linewidth]{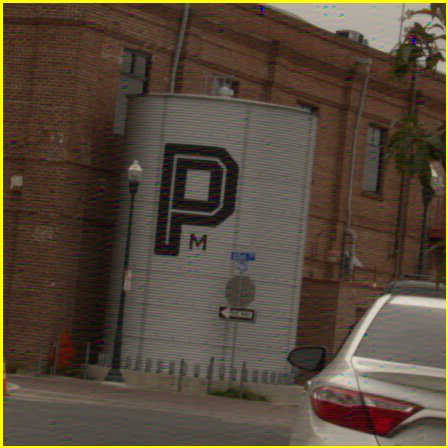} &
        \includegraphics[width=0.1\linewidth]{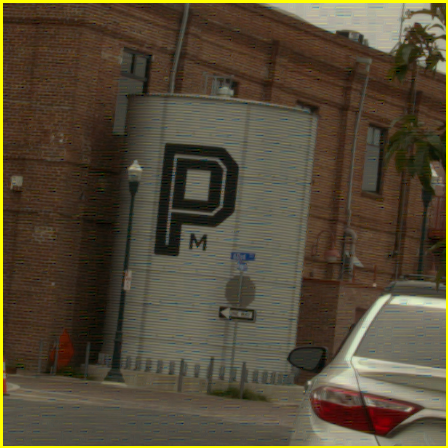} &
        \includegraphics[width=0.1\linewidth]{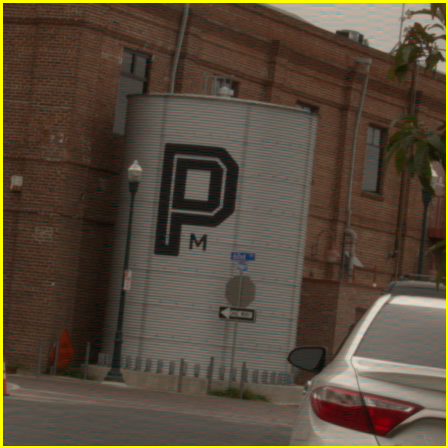} &
        \includegraphics[width=0.1\linewidth]{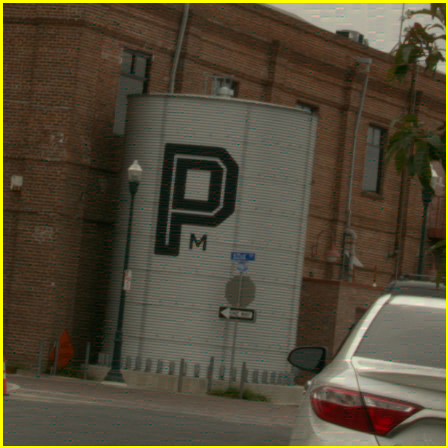} &
        \includegraphics[width=0.1\linewidth]{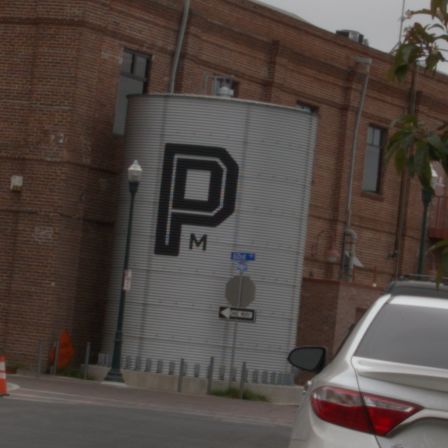} \\         

        \multirow{2}{*}[4.3em]{\includegraphics[width=0.21\linewidth]{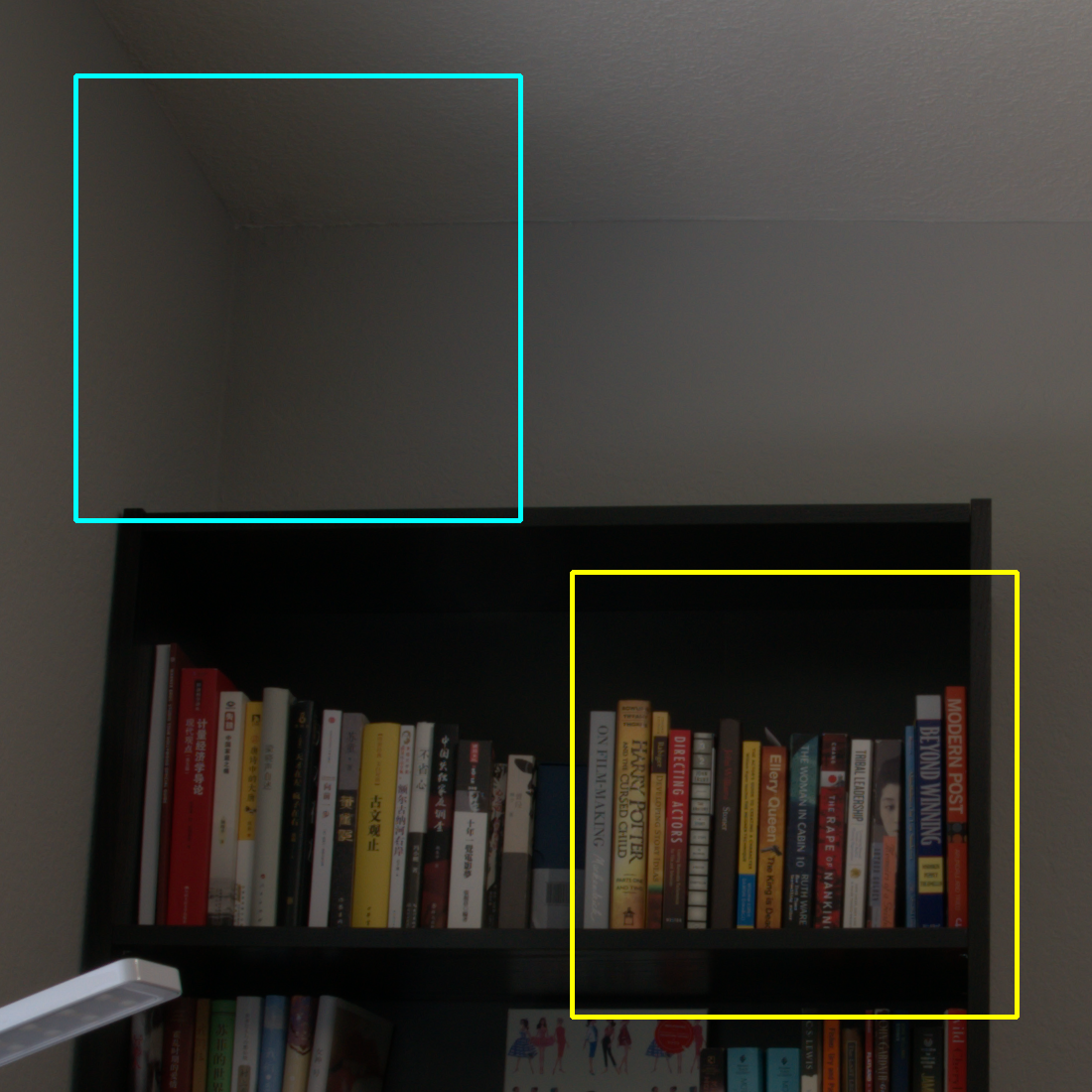}} & 
        \includegraphics[width=0.1\linewidth]{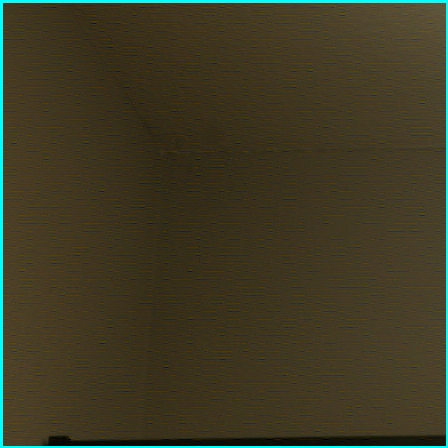} &
        \includegraphics[width=0.1\linewidth]{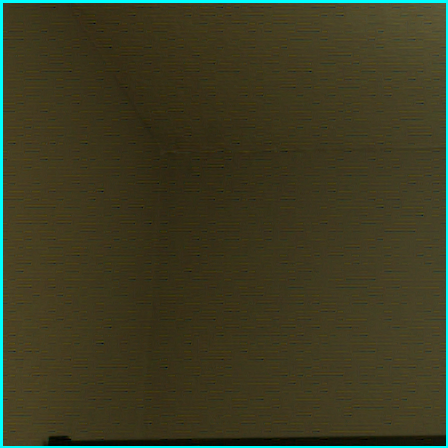} &
        \includegraphics[width=0.1\linewidth]{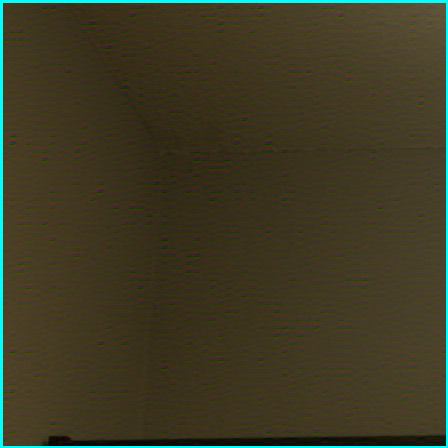} &
        \includegraphics[width=0.1\linewidth]{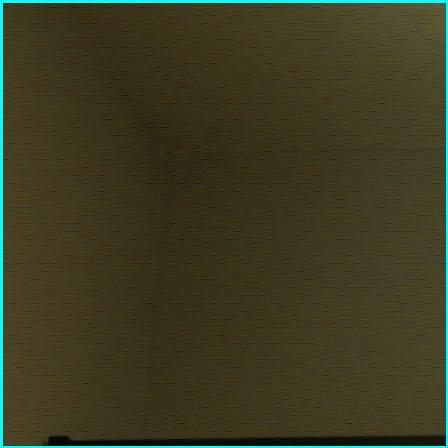} &
        \includegraphics[width=0.1\linewidth]{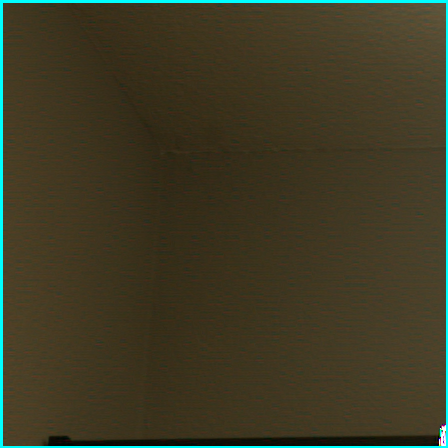} &
        \includegraphics[width=0.1\linewidth]{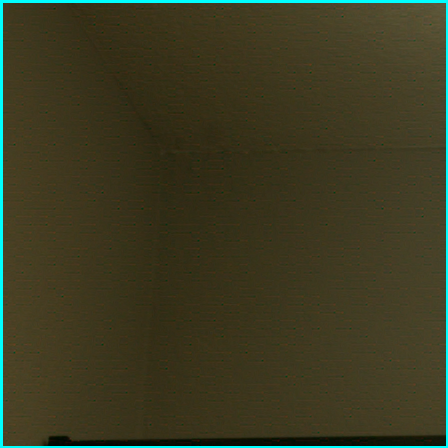} &
        \includegraphics[width=0.1\linewidth]{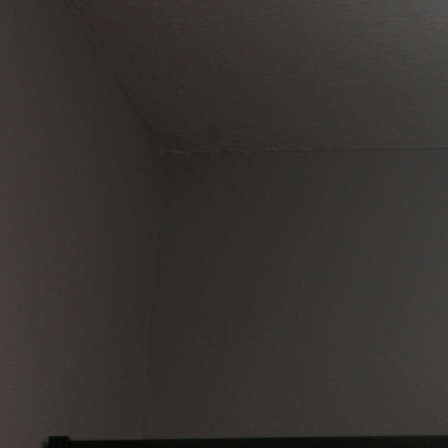} \\ 

        & 
        \includegraphics[width=0.1\linewidth]{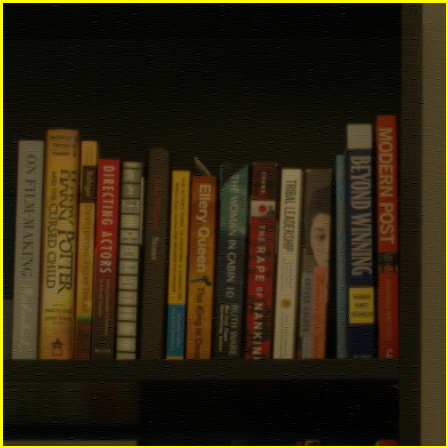} &
        \includegraphics[width=0.1\linewidth]{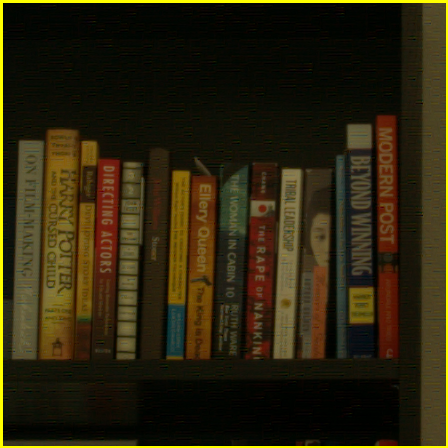} &
        \includegraphics[width=0.1\linewidth]{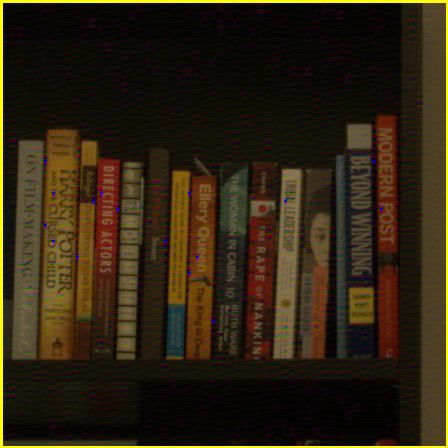} &
        \includegraphics[width=0.1\linewidth]{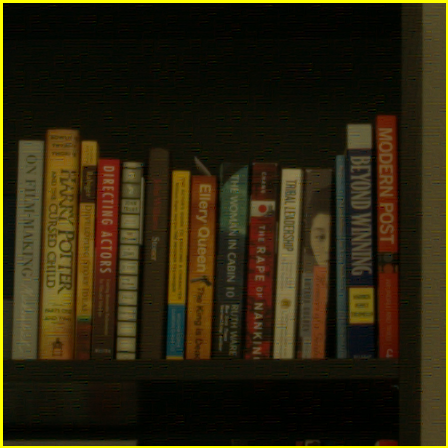} &
        \includegraphics[width=0.1\linewidth]{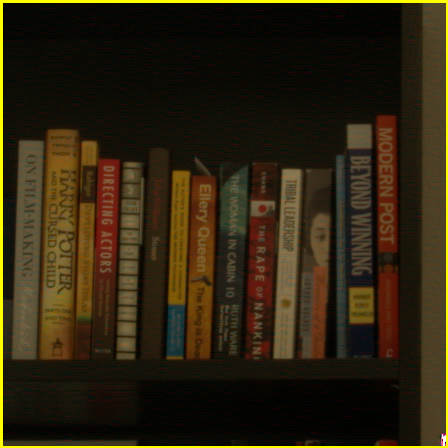} &
        \includegraphics[width=0.1\linewidth]{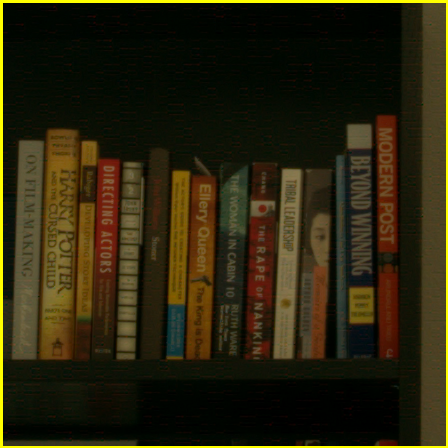} &
        \includegraphics[width=0.1\linewidth]{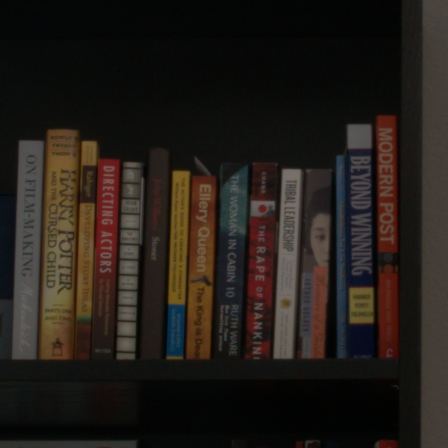} \\    

        \multirow{2}{*}[4.8em]{\includegraphics[width=0.21\linewidth]{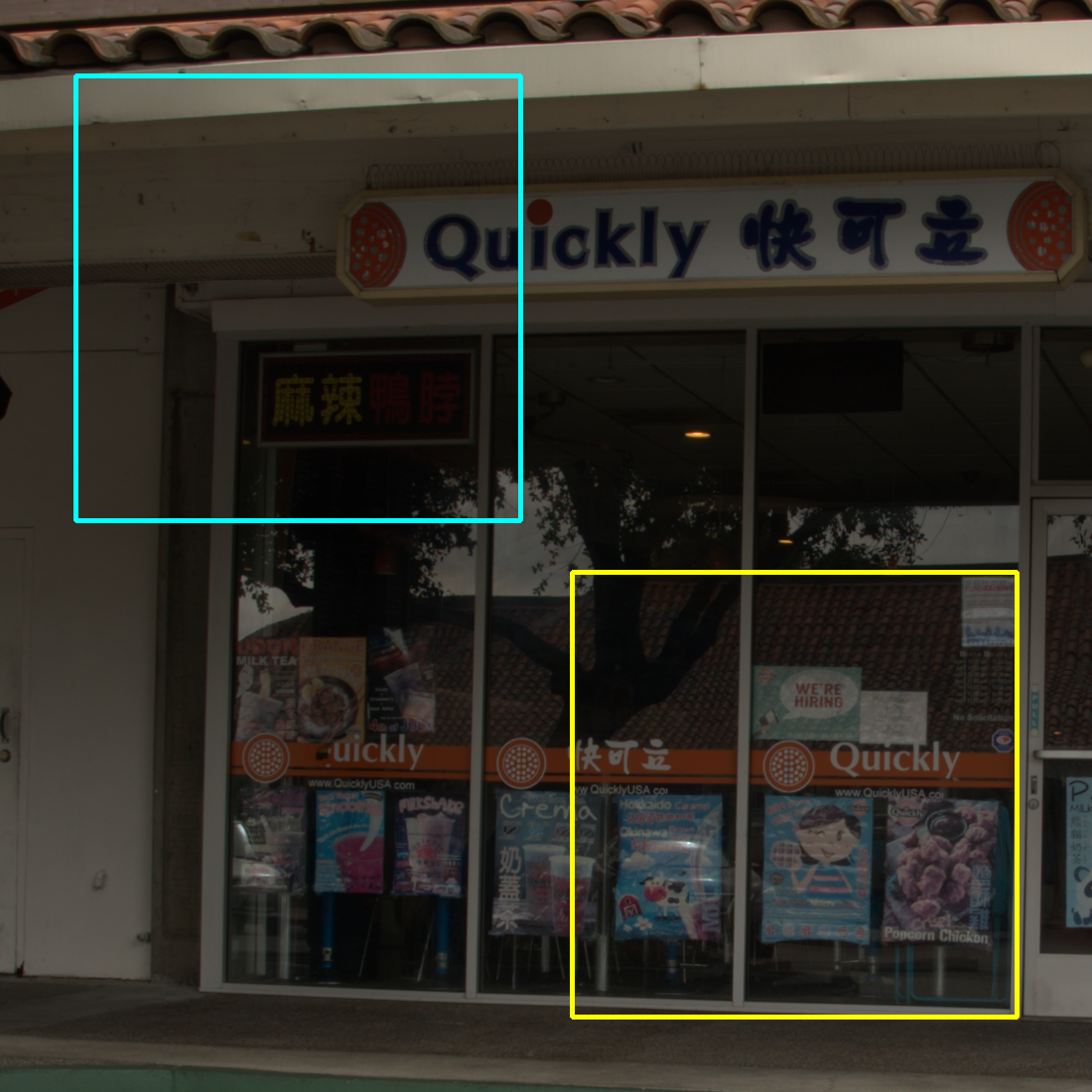}} &
        \includegraphics[width=0.1\linewidth]{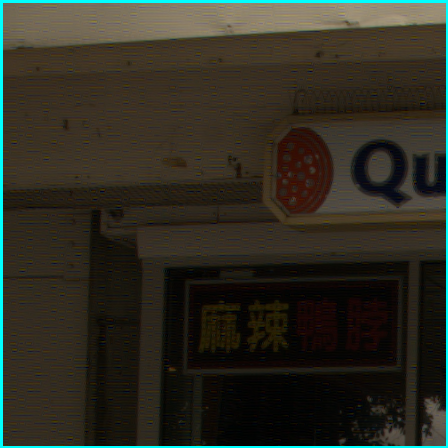} &
        \includegraphics[width=0.1\linewidth]{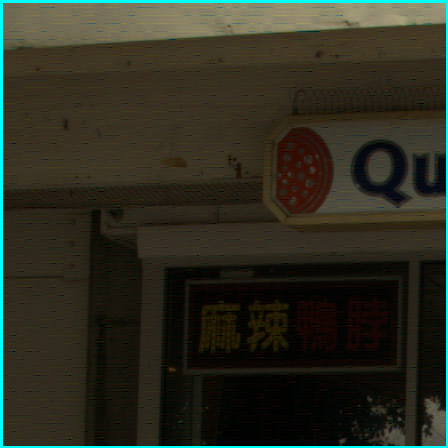} &
        \includegraphics[width=0.1\linewidth]{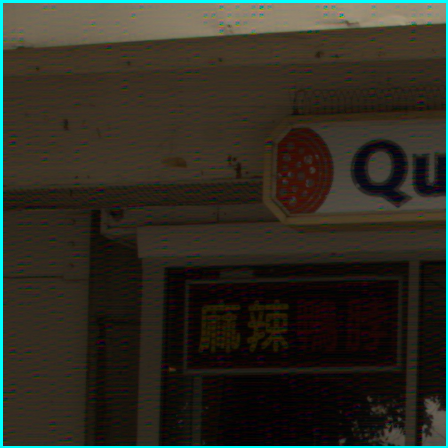} &
        \includegraphics[width=0.1\linewidth]{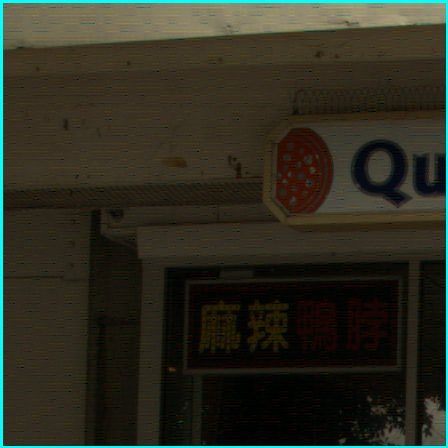} &
        \includegraphics[width=0.1\linewidth]{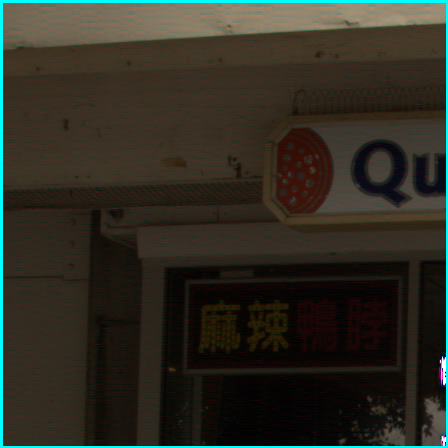} &
        \includegraphics[width=0.1\linewidth]{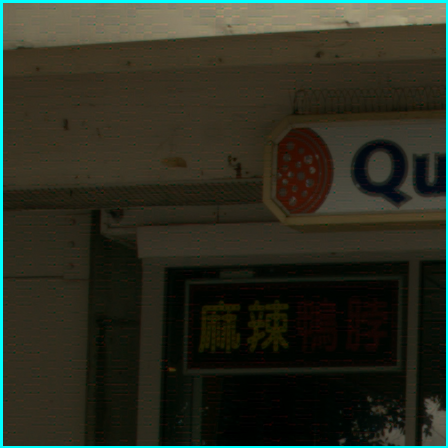} &
        \includegraphics[width=0.1\linewidth]{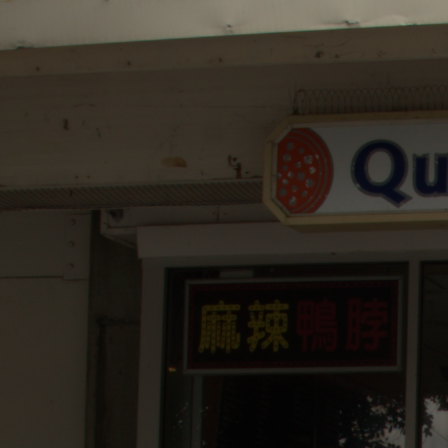} \\ 
        & 
        \includegraphics[width=0.1\linewidth]{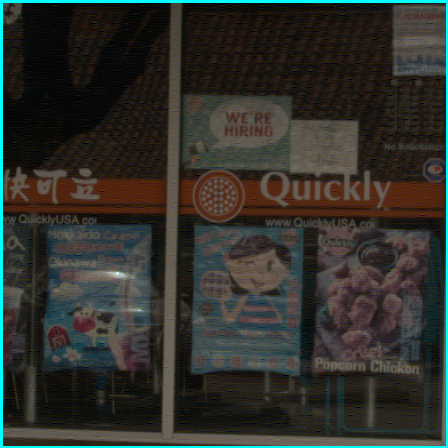} &
        \includegraphics[width=0.1\linewidth]{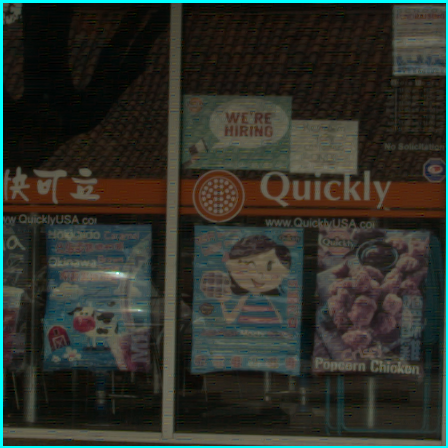} &
        \includegraphics[width=0.1\linewidth]{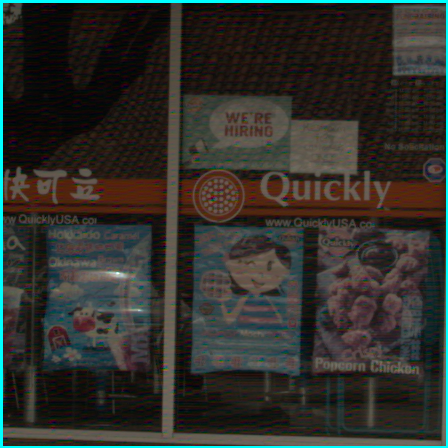} &
        \includegraphics[width=0.1\linewidth]{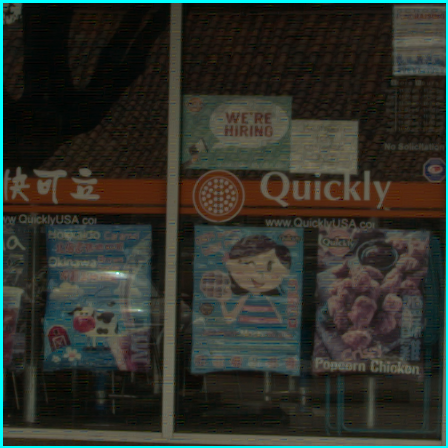} &
        \includegraphics[width=0.1\linewidth]{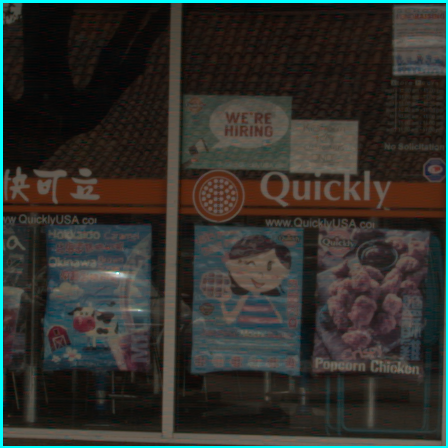} &
        \includegraphics[width=0.1\linewidth]{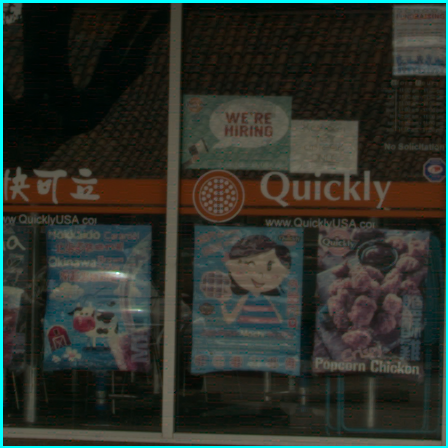} &
        \includegraphics[width=0.1\linewidth]{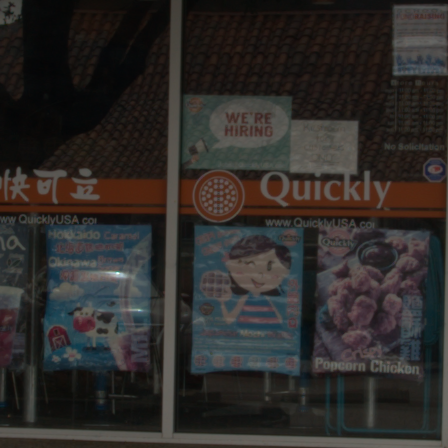} \\

        (a) RGB Images & (b) A w/o & (c) A w & (d) M w/o & (e) M w & (f) U w/o & (g) U w & (h) GT
    \end{tabular}
  \caption{
  % \textbf{Visual results on the SR-RAW test set. Training with distortion network will reduce the visual quality of the resulting RGB images but improve the robustness of watermarking. (b-g) [First key: A: AWNet, M: MW-ISPNet, U: UNet; Second key: C: Combined encoder; Third key: W: training with distortion network, O: training without distortion network.]}
  % }
  \textbf{Visual results on the SR-RAW test set. Training with distortion network will reduce the visual quality of the resulting RGB images but improve the robustness of watermarking. The combined encoder is used in the training of the above models. AWNet, MW-ISPNET, and UNet are abbreviated as A, M, and U.}
  }
  \label{figisp}
\end{figure*}

%% file: Contents/Limitations.tex
\section{Limitations}
\label{sec:Limitations}
Since our method uses the deep ISP pipeline, the training speed will slower than RGB watermarking methods. This is because the gradient of the entire ISP pipeline needs to be calculated. In the future, using a smaller deep ISP pipeline can improve the training speed.
At present, our method is limited to encoding and decoding RAW image pixel blocks of a fixed size. However, due to the considerable size of the RAW image in practical applications, we have opted to encode only the central area pixel block. If we choose other pixel blocks to add watermarking, we need to manually locate the watermarking pixel block from the RGB image. At the same time, since we use pixel blocks to add watermarking, the whole RGB results will have an obvious boundary around watermarked blocks.

%% file: Contents/Conclusion.tex
\section{Conclusion}
\label{sec:Conclusion}
This paper introduces an innovative end-to-end framework, referred to as \textbf{RAWIW}, with the primary goal of providing copyright protection for RAW images. The framework consists of a watermarking module, a deep ISP module, and a distortion network. Additionally, we employ an effective three-stage training strategy to strike a balance between the robustness and concealment of watermarking. The proposed method is validated on two RAW image datasets, namely ZRR dataset and SR-RAW dataset. The results demonstrate that the RAWIW framework exhibits robustness to different ISP pipelines and distortions during transmission, all while minimizing the impact on the visual quality of the resulting RGB images. As for future work, we plan to delve further into investigating strategies to enhance the robustness and concealment capabilities of the watermarking process. By continuously refining the framework, we aim to bolster its performance and broaden its applicability in various scenarios related to RAW image copyright protection.